\documentclass[12pt,preprint]{aastex}

\newcommand{\ngal}{N$_{gal}$}
\newcommand{\ngrtwo}{N$_{gal}^{R200}$}

\newcommand{\rtwo}{R$_{200}$}

\newcommand{\msun}{M$_\sun$}
\newcommand{\lsun}{L$_\sun$}
\newcommand{\up}{$u$}
\newcommand{\gp}{$g$}
\newcommand{\rp}{$r$}
\newcommand{\ip}{$i$}
\newcommand{\zp}{$z$}
\newcommand{\gmr}{$g-r$}

\newcommand{\hinv}{$h^{-1}\ $}

\newcommand{\libcg}{$L_{i}^{BCG}$}
\newcommand{\limem}{$L_{i}^{mem}$}
\newcommand{\lrbcg}{$L_{r}^{BCG}$}
\newcommand{\lrmem}{$L_{r}^{mem}$}

\newcommand{\lstar}{L$_{*}$}
\newcommand{\dsym}{$^{\circ}$}

\shorttitle{MaxBCG Cluster Catalog}
\shortauthors{Koester et al.}

\begin{document}
\title{A MaxBCG Catalog of 13,823 Galaxy Clusters from the Sloan Digital Sky Survey}
\author{B. P. Koester\altaffilmark{1},\email{bkoester@umich.edu} T. A. McKay\altaffilmark{1,2},\email{tamckay@umich.edu}
J. Annis \altaffilmark{3}, R. H. Wechsler\altaffilmark{4}, 
A. Evrard\altaffilmark{1,2}, L. Bleem\altaffilmark{5}, 
M. Becker\altaffilmark{1}, D. Johnston\altaffilmark{6}, 
E. Sheldon\altaffilmark{7}, R. Nichol\altaffilmark{8}, 
C. Miller\altaffilmark{9}, R. Scranton\altaffilmark{10}, 
N. Bahcall\altaffilmark{11}, J. Barentine\altaffilmark{12}, 
H. Brewington\altaffilmark{12}, J. Brinkmann\altaffilmark{12}, 
M. Harvanek\altaffilmark{12}, S. Kleinman\altaffilmark{12}, 
J. Krzesinski\altaffilmark{12,13}, D. Long\altaffilmark{12}, 
A. Nitta\altaffilmark{12}, D. Schneider\altaffilmark{14}, 
S. Sneddin\altaffilmark{12}, W. Voges\altaffilmark{15}, 
D. York\altaffilmark{16}, SDSS collaboration\altaffilmark{17}}


\altaffiltext{1}{Department of Physics, University of Michigan,
    Ann Arbor, MI 48109}
\altaffiltext{2}{Michigan Center for Theoretical Physics}
\altaffiltext{3}{Fermi National Accelerator Laboratory, Batavia, IL, 60510}
\altaffiltext{4}{Kavli Institute for Cosmological Physics, Department of Astronomy and
Astrophysics, and Enrico Fermi Institute, University of 
  Chicago, Chicago, IL, 60637; Hubble Fellow}
\altaffiltext{5}{Department of Physics, University of Chicago, Chicago, IL, 60637}
\altaffiltext{6}{California Institute of Technology, Jet Propulsion Laboratory, 1200 E. California, Pasadena, CA, 91106}
\altaffiltext{7}{Center for Cosmology and Particle Physics, New York University, New York, NY 10003}
\altaffiltext{8}{Institute of Cosmology and Gravitation, Mercantile House, Hampshire Terrace, University of Portsmouth, Portsmouth P01 2EG, UK}
\altaffiltext{9}{Cerro-Tololo Inter-American Observatory, NOAO, Casilla 603, LaSerena, Chile}
\altaffiltext{10}{University of Pittsburgh, Department of Physics and Astronomy, 3941 O'Hara Street, Pittsburgh, PA 15260}
\altaffiltext{11}{Princeton University Observatory, Peyton Hall, Princeton, NJ 08544}
\altaffiltext{12}{Apache Point Observatory, P.O. Box 59, Sunspot, NM 88349}
\altaffiltext{13}{Mt. Suhora Observatory, Cracow Pedagogical University, ul. Podchorazych 2, 30-084 Cracow, Poland}
\altaffiltext{14}{Department of Astronomy and Astrophysics, 525 Davey Laboratory, Pennsylvania State University, University Park, PA 16802}
\altaffiltext{15}{Max-Planck-Institute fur extraterrestriche Physik, Geissenbachstrasse 1, D-86741, Garching, Germany}
\altaffiltext{16}{Astronomy and Astrophysics Center and Enrico Fermi Institue, University of Chicago, 5640 South Ellis, Chicago, IL 60637}
\altaffiltext{17}{www.sdss.org}


\begin{abstract}
We present a catalog of galaxy clusters selected using the maxBCG red-sequence method from Sloan Digital Sky Survey photometric data. This catalog includes 13,823 clusters with velocity dispersions greater than $\approx$400 km~s$^{-1}$, and is the largest galaxy cluster catalog assembled to date. They are selected in an approximately volume-limited way from a 0.5 Gpc$^3$ region covering 7500 square degrees of sky between redshifts 0.1 and 0.3. Each of these clusters contains between 10 and 190 E/S0 ridgeline galaxies brighter than 0.4 \lstar\ within a scaled radius \rtwo. The tight relation between ridgeline color and redshift is used to determine photometric redshift estimates for every cluster. In addition, SDSS spectroscopic redshifts are available for at least the brightest galaxy in 39\% of the clusters. Tests of purity and completeness are obtained by running the cluster finder on realistic mock catalogs. These studies suggest that the sample is more than 90\% pure and more than 85\% complete for clusters with masses $\geq 1 \times 10^{14}$ solar masses. Photometric redshift errors are shown by comparison to spectroscopic redshifts to be small ($\Delta_{z} \simeq 0.01$), essentially independent of redshift, and well determined throughout the redshift range. Spectroscopic measurements of cluster members are used to examine the extent to which projection along the line of sight contaminates identification of brightest cluster galaxies and cluster member galaxies. Spectroscopic data are also used to demonstrate the correlation between optical richness and velocity dispersion. Comparison to the combined NORAS and REFLEX X--ray selected cluster catalogs shows that X--ray luminous clusters are found among the optically-richer maxBCG clusters. This paper is the first in a series which will consider the properties of these clusters, their galaxy populations, and their implications for cosmology.
\end{abstract}



\section{Introduction}

Galaxy clusters are the most visible
features of large scale structure. They occupy very massive dark
matter halos, and are observationally accessible by many means. Their locations are
revealed by the presence of large numbers of tightly clustered
galaxies, pools of hot X--ray emitting gas, and relatively strong
features in the gravitational lensing shear field. Many of their
properties, including their number as a function of dark matter mass and their
spatial clustering, can be predicted with confidence from N-body
simulations. Since clusters occupy the tail of the halo mass function,
their numbers are exponentially sensitive to variations in
cosmology. Precise observations of large numbers of clusters provide
an important tool for testing our understanding of cosmology and
structure formation. Clusters are also interesting laboratories for
the study of galaxy evolution under the influence of extreme
environments.

Clusters were first detected in the 18th century \citep{biv00} as
significant overdensities of galaxies on the sky. The early 20th
century saw the first explorations of their physical properties, 
including the discovery of dark matter in Zwicky's study of
Coma \citep{zwi33, zwi37}. Statistical studies of the cluster
population became possible with the introduction of large area surveys
implementing uniform selection methods. For example, Abell and his 
collaborators \citep{abe58, abe89} created a sample of 4073 clusters
over the whole sky by visually inspecting photographic plates.
 They identified clusters as overdensities of 30 or more galaxies (each no
more than two magnitudes fainter than the third brightest member,
${\rm m_3}$) within a fixed metric aperture on the sky. Distances
were estimated based on the magnitude of the tenth brightest
cluster member. Similar catalogs, with somewhat different selection
criteria, were developed by \citet{zwi61}.

The Abell catalog has been very influential \citep[see, e.g.,][ and
references therein]{bah99}, supporting
 a wide range of cluster
science. Among its most important achievements, it enabled 
 some of
the first studies of the large-scale distribution of matter in the 
Universe \citep{bah83, pos92, mil99}. The brightest galaxies in
clusters have also
 been used to construct Hubble diagrams
\citep{gun75, san76, pos95}. 
 Studies of the Abell cluster catalogs
have been hampered by problems of projection \citep{sut88, col95},
false clusters \citep{luc83}, and an uncertain selection function.  
Projection refers to the unwanted effect of unassociated foreground
and background galaxies on the measured properties of clusters (richness, 
${\rm m_3}$, etc.), while `false clusters' are objects overdense in projection, 
but not physically bound. Uncertainties in the selection function arise from the 
use of visual selection and the relatively poor calibration of photographic data.

These problems prompted the construction of new optically--selected samples
of clusters using digitized photographic material and automated
cluster selection algorithms \citep{lum92, dal94, gal03}. In recent
years, 
 the increased availability of large-area CCD photometry has
prompted 
 a new round of optical cluster selection \citep[e.g.,][
and references therein]{gal06}. For example, \citet{pos96} 
constructed a sample of clusters using a ``matched filter'' technique,
smoothing the galaxy distribution with a filter optimized for the
detection of distant clusters. Another example is the direct search
for clusters as resolved sources
 of diffuse optical light
\citep{dal96, zar97, gon01}.
 
 The advent of multi--color CCD photometry obtained over large areas has had a major impact. 
Precise measurement of galaxy color has
literally added a new dimension to optical cluster finding. Clusters
are dominated by old, red E/S0 galaxies which occupy a narrow region
in color-magnitude space known as the E/S0 ridgeline
\citep{bow92}. This tight clustering of galaxies in color,
magnitude, and space allows significant improvements in cluster 
finding \citep{ost98, gla00, bah03, gla05}. The location 
 of this
ridgeline in color shifts smoothly with redshift, providing quite
precise 
 estimates of cluster redshift. These new ``red sequence''
techniques have largely eliminated projection effects and false
clusters from optical cluster catalogs, making it possible to perform
accurate measurements of the large-scale structure in the universe
using such samples.

In parallel with advances in optical searches for clusters, there has
been considerable progress in cluster identification in X--rays.
Over the last decade, a variety of cluster catalogs have
been constructed from X--ray surveys of the sky \citep{gio94, ebe96, 
ros98, rom00, boe00, mul03, boe04}. These surveys are less sensitive 
to projection effects, though they can be contaminated by AGN emission.
In addition, the relation of X--ray luminosity and temperature to the 
underlying total mass of a cluster is thought to be more accessible to
theoretical prediction than optical mass proxies. 
The principle limitation of X--ray cluster surveys is the
relative scarcity of the observing resource. Existing all-sky surveys
are limited to detection of rare, high-flux sources. As a result, the
total number of groups and clusters yet detected in X--rays is modest;
there are only 1579 objects in the BAX 
database, a compendium of all X--ray groups and clusters detected as of 2004
\citep{sad04}. With no future all--sky X--ray survey planned, it is 
unlikely the number of X--ray detected clusters will increase 
substantially over the next decade, though important contributions will be made
by serendipitous surveys in the XMM and Chandra archives \citep{rom01, gre04, rom04}.

Clusters have also been detected as peaks in the shear field of deep weak lensing
surveys \citep{erb00, clo01, mir02, wit01, dah03, wit03, wit05}. This method has the 
virtue of directly probing
the projected surface mass density of the clusters. Unfortunately, this projection
introduces substantial noise in the mass estimation for individual clusters \citep{whi02, dep05}.
While lensing provides a unique test for the presence of truly dark clusters, it
is otherwise an expensive detection method. To detect a cluster using lensing,
images must be obtained of a large number of faint galaxies in the background of a 
cluster. The same images used to detect the shear always contain very high signal-to-noise
detections of many cluster galaxies, galaxies which would have been easily 
detected in much shallower imaging. While lensing is probably not an optimal 
method for cluster detection, is is a vital tool in cluster mass 
calibration \citep{she01, jon05, she06} for clusters detected by any other method.

In the future, large cluster catalogs will be constructed using
observations of the Sunyaev-Zeldovich (SZ) effect. The X--ray
emitting hot gas in clusters causes Compton up-scattering of a small
fraction of the CMB photons which pass through it. This leads to a characteristic distortion in
the transmitted CMB spectrum \citep{car02}. Observation of these
distortions provides an additional method for cluster detection. The
SZ effect nicely combines the guarantee of a deep potential well provided by
the hot gas along with a detection signature which is essentially independent of
redshift. Of course SZ surveys must
be supplemented by follow-up optical observations to determine redshifts if they
are to be useful for cosmology. They will also benefit from the mass calibrations
which weak lensing measurements provide. As a result, combined SZ and optical surveys
are a particularly promising approach for cluster study in the coming years.

The most significant challenge in cluster science lies in relating
what is best understood about clusters from theory to what is measured
in observations. Structure formation simulations can predict the
evolution of the dark matter component of clusters with great
confidence \citep[e.g.,][]{eve02}. Observations provide measures of
the cluster galaxies and gas with ever-increasing precision.
Unfortunately, the cluster baryons we observe participate in
hydrodynamic and thermodynamic interactions which are too complex for
current simulations to confidently model.  This leaves an
uncomfortably uncertain gap between theory and observation. Closing
this gap is the goal of a substantial body of current research. New
simulations of many kinds are pushing the ability to predict the
evolution of galaxies and gas \citep[e.g.,][]{spr05, kra06} while new
observations are extracting ever more direct and precise measurements
of total cluster mass proxies \citep{vik06, dia05, maj04, kat04}.

This paper describes a large new catalog of galaxy clusters extracted
from Sloan Digital Sky Survey optical imaging data (SDSS: \citealt{yor00}). It is the first in
a series of papers which will explore the properties of optically-selected SDSS cluster samples 
and their use for cosmology. Galaxy clusters are selected
by calculating the likelihood that each observed galaxy is the
brightest galaxy in a cluster. This likelihood is based on the galaxy's 
color and magnitude, along
with the degree to which other galaxies are clustered around it in color,
magnitude, and space. The resulting catalog is an essentially volume-limited list of galaxy group and cluster locations together with
estimates of their total galaxy content and redshift. We present here
a catalog of the richest objects detected by this method, including
13,823 clusters, each containing ten or more E/S0 ridgeline galaxies
brighter than 0.4 L* (in the $i$-band) within a scaled radius \rtwo,
 drawn from about 7500 square degrees of sky, and
extending over a redshift range from z=0.1-0.3.

Section \ref{data} presents a brief description of the SDSS data
relevant for this paper. This is followed in Section \ref{algorithm}
by an outline of the cluster selection algorithm, details of which are
presented in a companion paper \citep{koe06}. Overall properties of
the derived cluster catalog are presented in Section \ref{catalog}.
Basic tests of the completeness and purity of cluster selection based
on mock catalogs are described in Section \ref{mocks}. Tests of the
quality of redshift estimates, the importance of projection effects,
the relationship between cluster richness measures and velocity dispersion, and the
space density of clusters are outlined in Section \ref{tests}. As an
additional test, comparison of this catalog to existing X--ray selected
cluster catalogs is provided in Section \ref{X--ray comparison}. We
conclude in Section \ref{discussion}. Where needed, we assume a standard 
$\Lambda$CDM cosmology with $\Omega_M$ = 0.3, $\Omega_{\Lambda}$ = 0.7, and 
H$_0$ = 70 km s$^{-1}$ Mpc$^{-1}$.

\section{Data \label{data}}

Data for this study are drawn from Sloan Digital Sky Survey\footnote{www.sdss.org}: a combined imaging and spectroscopic survey of 10$^4$ deg$^2$ in the North Galactic Cap and a smaller region in the South. The imaging survey was carried out using a specially designed 2.5 m telescope \citep{gun06} in drift-scan mode in five SDSS filters (\up, \gp, \rp, \ip, \zp) to a limiting magnitude of r$<$22.5 \citep{fuk96, gun98, lup99, hog01, smi02}. Photometric errors are typically limited at bright magnitudes by systematic uncertainties at the $\le$3\% level \citep{ive04}. Astrometric errors are typically smaller than 50 mas per coordinate \citep{pie03}. The spectroscopic survey targets both a `main' sample of galaxies with \rp$<$17.8 and a median redshift of z$\sim$0.1 \citep{str02, bla03} and a `luminous red galaxy' sample \citep{eis01} which is approximately volume limited out to z=0.38. For more details of early and more recent SDSS data releases see \citet{sto02} and \citet{ade06}. 

\begin{figure}
\begin{center}
\rotatebox{90}{\scalebox{0.7}{\plotone{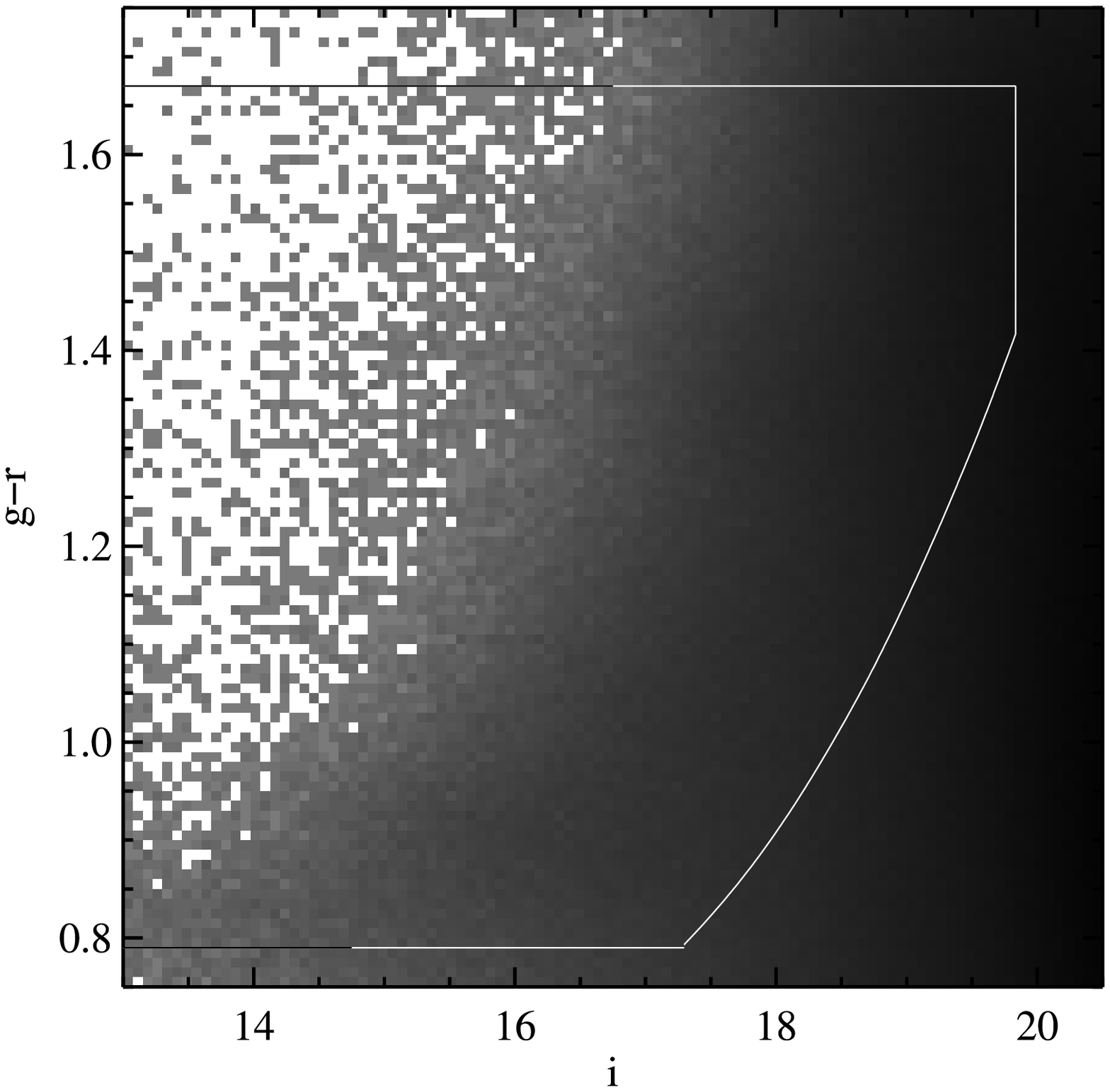}}}
\figcaption{The distribution of galaxies in \gmr\ color and \ip\ magnitude in the input galaxy catalog is displayed in this figure. Overlaid on the figure are the cuts used to extract a sample of potential E/S0 ridgeline galaxies in the search redshift region. The gray scale is linear in the number of galaxies at each color and magnitude. \label{input catalog}}
\end{center}
\end{figure}

Two catalogs are extracted from the SDSS data for use in this paper. The first is a photometric galaxy catalog, used as input to the cluster finder. This catalog is generated using the same criteria applied in \citet{scr02} to create the input catalog for measurements of galaxy clustering, including use of an optimized star--galaxy separator. The input catalog includes galaxy positions and `CMODEL' magnitudes in each of the SDSS bands. These magnitudes are constructed from a weighted combination of Petrosian and model magnitudes, with the weights determined by the quality of individual fits of deVaucoleurs and exponential profiles to the surface brightness profile of each galaxy. All magnitudes are corrected for Galactic extinction using the extinction maps of \citet{sch98}. The resulting input catalog contains $2.3 \times 10^{7}$ galaxies. The distribution of these galaxies in \gmr\ color and \ip\ magnitude is shown in Figure \ref{input catalog}.

The galaxies in our imaging catalogs include a wide range of objects with various colors, magnitudes, and morphologies. Only a subset of these have colors and magnitudes consistent with E/S0 ridgeline galaxies in the target redshift range. Imposing broad color cuts to exclude galaxies too red, blue, or faint to be bright cluster members substantially reduces the size of the input catalog with no loss in cluster finding efficiency. The colors and magnitudes appropriate for E/S0 galaxies in the redshift range $0.1 < z < 0.3 $ are extracted using the color-magnitude-redshift model described in \citet{koe06}. We focus on this redshift range for reasons explained in more detail in Section \ref{catalog}. These model parameters are turned into color-magnitude cuts in the following way: at an assumed redshift, we note the $g-r$ and $r-i$ colors and 0.4 \lstar\ $i$-band magnitude prescribed by the model. We include in the search catalog all galaxies whose colors $(g-r)_{candidate}$ and $(r-i)_{candidate} $ pass the following criteria anywhere in the search redshift range:

\begin{eqnarray}
& (g-r)(z)_{model}-\sqrt{\sigma_{err}^2+0.15^2} < (g-r)_{candidate} < (g-r)(z)_{model}+\sqrt{\sigma_{err}^2+0.15^2}  \\
& (r-i)(z)_{model}-\sqrt{\sigma_{err}^2+0.18^2} < (r-i)_{candidate} < (r-i)(z)_{model}+\sqrt{\sigma_{err}^2+0.18^2}  \nonumber \\
& i_{candidate} < 0.4\ L_*(z)^{model} \nonumber
\end{eqnarray}

\noindent Here, the $\sigma_{err}^2$ are the color errors of the objects, measured using the methods described in \citet{scr05}. The factors of 0.15 and 0.18 each correspond to 3 times the intrinsic width of the E/S0 ridgeline in $g-r$ and $r-i$, respectively. At the same redshift, a magnitude constraint selects objects whose $i$-band magnitudes indicate that they are 0.4 \lstar\ or brighter at the current redshift. Running these selection criteria for the chosen redshift range reduces the total number of input galaxies to 4,689,495, or about one fifth of the input galaxy catalog (see Figure \ref{input catalog} for the range of these cuts in $g-r$ vs. $i$, excluding color errors). These galaxies are all treated as potential centers of groups or clusters by the maxBCG cluster finder described below. We note that this will bias the cluster search against completely blue, low redshift clusters and groups (Section 3). Most of these input galaxies (a total of 1,389,858) are ultimately absorbed as centers or members of maxBCG groups and clusters.

\begin{figure}
\rotatebox{0}{\scalebox{0.9}{\plotone{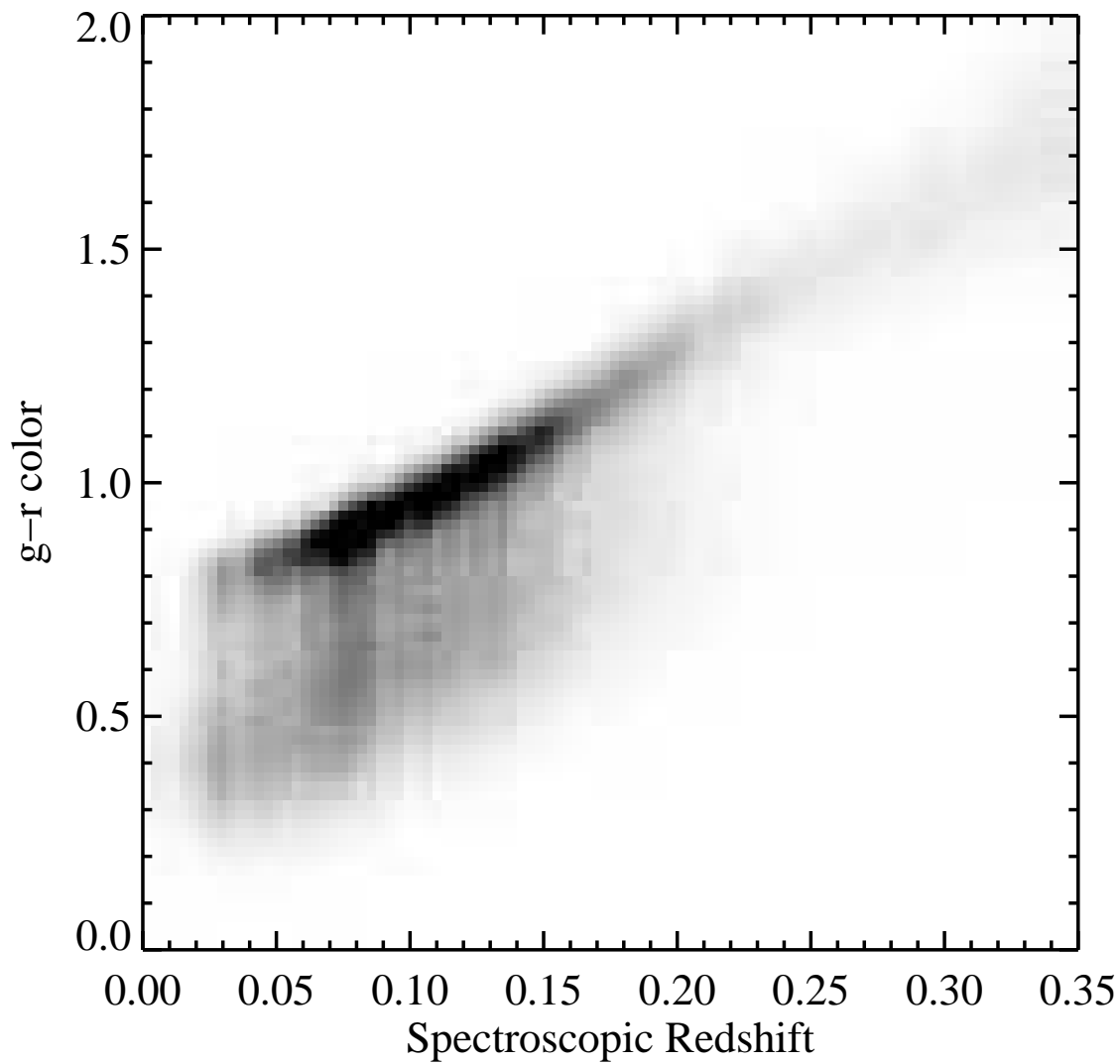}}}
\figcaption{This figure shows the distribution of galaxies in the input spectroscopic catalog in \gmr\ color and redshift. The close correspondence between redshift and \gmr\ color for the E/S0 galaxies is apparent in the strong sequence along the red edge of the distribution. This forms the basis for the excellent photometric redshifts obtained for maxBCG clusters. The gray scale is linear in the number of galaxies at each color and redshift. \label{input spectroscopic}}
\end{figure}

The second input catalog is drawn from the SDSS spectroscopic data, including all available spectroscopic targets identified as galaxies with confidently measured redshifts. The catalog used here contains 567,486 galaxies with typical radial velocity errors of about $\pm$30 km~s$^{-1}$. This spectroscopic catalog is not used in the detection of clusters, but provides vital data for testing the fidelity of photometric redshift estimation. Spectroscopic data also allows us to determine the relationship between measured cluster richness and velocity dispersion. The \gmr\ color as a function of redshift for this spectroscopic catalog is shown in Figure \ref{input spectroscopic}. 

\section{Description of the Cluster Detection Algorithm \label{algorithm}}

Galaxy cluster detection within the SDSS has been presented in several previous works. Initial efforts at photometric selection of clusters were presented in \citet{got02} and \citet{bah03}. The latter compared the results of two different selection methods, a hybrid matched filter method \citep{kim02} and an earlier version of the ``maxBCG'' method presented here \citep{ann99}. A photometrically selected catalog of compact groups has also been assembled \citep{lee04}. In addition to these photometrically selected catalogs, several group and cluster catalogs have been assembled from spectroscopic data, for example \citet{mil05}, \citet{ber06}, and \citet{wei06}. These spectroscopic catalogs are quite robust, but limited in volume by the flux-limited nature of the SDSS spectroscopy. Previously detected clusters have also been studied using SDSS optical data. Popesso and collaborators have published a series of papers relating X--ray and optical properties of Abell clusters \citep{pop04, pop05, pop06}. Studies have also been conducted of the galaxy populations \citep{han05} and lensing masses \citep{she01} of an earlier generation of maxBCG objects. Luminosity functions for photometrically selected clusters were considered in \citet{got03}.

The maxBCG galaxy cluster selection algorithm applied to the input galaxy catalog is described in more detail in a companion paper \citep{koe06}.  Briefly, the algorithm exploits two well-known features of rich galaxy clusters. First, the bright end of the cluster luminosity function is dominated by galaxies occupying a narrow region of color-magnitude space (the E/S0 ridgeline). These galaxies are sometimes referred to as red-sequence galaxies. Second, clusters contain a brightest cluster galaxy (BCG) that is located near the center of the galaxy distribution. This BCG is often distinctly brighter than other cluster members \citep{han05, loh06}, and nearly at rest relative to the cluster center of mass \citep{oeg01}. While these criteria are not universal for all groups and clusters, they become increasingly common among the optically richest clusters.

For every galaxy in the input catalog, the algorithm measures two independent likelihoods. The first is the likelihood that a galaxy is spatially located in an overdensity of E/S0 ridgeline galaxies with similar g-r and r-i colors, and the second is the likelihood that it has the color and magnitude properties of a typical BCG. Both likelihoods are evaluated for every input SDSS galaxy at a grid of redshifts. The redshift which maximizes the product of these likelihoods is then found for each galaxy. This corresponding maximum likelihood redshift provides a first estimate of the cluster redshift, and is used in the following steps. 

Next a list of member galaxies for each potential center is assembled. The number of galaxies projected within 1 \hinv Mpc of this potential center, brighter than 0.4 \lstar, fainter than the potential center, and with colors matching its E/S0 ridgeline, is counted. This number, which we call \ngal, provides a first estimate of cluster richness. This initial richness estimate is then used to estimate cluster size \rtwo, defined here as the radius within which the density of galaxies with $-24 \le M_r \le -16$ is 200 times the mean density of such galaxies. This \rtwo\ is determined using the relation between \ngal\ and \rtwo\ derived in \citet{han05}. This richness-size relation was determined for a somewhat different, earlier version of the maxBCG cluster finder. While the normalization of the richness-size relation may differ slightly in this new catalog, we expect the overall scaling to remain the same. 

Once the list of cluster center likelihoods is measured and lists of corresponding potential members are assembled, galaxy clusters are assembled beginning with the richest and progressing down through a percolation procedure. The percolation begins with the highest likelihood potential center. This galaxy is declared a cluster BCG and assigned its maximum likelihood redshift. All galaxies within a projected separation \rtwo\ of the BCG, within $\pm 2\sigma$ of the E/S0 ridgeline in the space of \gmr\ and i magnitude, and brighter than 0.4 \lstar\ in i band are labeled as members of this cluster. Subsequent galaxies can be considered BCGs only if they are not already members of a higher likelihood center. We further eliminate any lower likelihood centers which fall within \rtwo\ and have a maximum likelihood redshift within $\pm$0.02 of a higher likelihood BCG. This latter step prevents the double counting which might occur if a real member of rich cluster has slightly anomalous colors which prevent it from being selected as a cluster member. Note that this local suppression has important implications for measurement of the spatial clustering of clusters. The observed clustering will be strongly affected by the cluster finding algorithm on scales less than a few Mpc. This process continues until all viable centers are either declared centers or otherwise absorbed as cluster members. Each object detected has a center defined as the BCG location, an estimated redshift, and a richness given by the number of E/S0 ridgeline members brighter than 0.4 L* and within \rtwo\ of the cluster center (\ngrtwo).  

Once the basic cluster finding step is complete, we refine the measurement of cluster properties in several ways. First, redshift estimates are adjusted. A small empirical correction ($\approx 0.004$) is applied to the photometric redshifts, based on the comparison of spectroscopic to photometric redshifts described below. Information about the spectroscopic redshift of the cluster, where it is available, is appended. The luminosity of the BCG and the summed luminosity of the BCG and all cluster members in r and i bands is added. Luminosities are reported in units of $10^{10} L_{\sun}$, k-corrected to z=0.25, the median redshift of the sample. K-corrections are computed using the LRG template in v4.1.4 of KCORRECT \citep{bla03b}, assuming the maxBCG photometric redshift.  These k-corrections do not include an evolution correction. The final cluster catalog contains an array of measured properties, including location, photometric redshift, spectroscopic redshift (where available), and several richness and mass estimators, including \ngal, \ngrtwo, \lrbcg, \libcg, \lrmem, and \limem. Once the cluster catalog has been constructed, many more refined measures of cluster properties, including overall richness, shape, concentration and galaxy content can be made. 

In concluding this description, it is worth noting that color cuts and the matched-filter employed in maxBCG restrict detections of completely blue groups and clusters, especially those found at low redshift \citep{got02}. Furthermore, it assigns lower likelihoods to objects with less well-defined ridgelines, unusual BCGs, or BCGs off-center from the cluster galaxy distribution. Although maxBCG may not miss such objects, they are of course penalized for their deviation from the model. More generally, since a perfect detection algorithm does not exist, understanding the selection function of any algorithm requires comparison to clusters detected by alternative approaches. In Section 7, X--ray selected clusters are compared to the maxBCG catalog, and simulations are used in Section 5 and also in \citep{koe06} to address the robustness of the detection algorithm.

\section{Overall Properties of the Derived Catalog \label{catalog}}

The algorithm described above produces a catalog of all groups and clusters from z=0.05-0.35 which contain bright red galaxies, a total of $2.18 \times 10^{6}$ objects. The resulting list spans a very broad mass range, from isolated ellipticals and small groups through rich clusters. We focus here on only the more substantial clusters: those containing at least 10 E/S0 ridgeline galaxies brighter than 0.4 \lstar\ within a scaled aperture \rtwo.  We focus on a restricted redshift range as well. SDSS photometric data are particularly well suited for the measurement of clusters at moderate redshift. In the redshift range from 0.1 to 0.3 the relation between redshift and \gmr\ color is particularly simple, reflecting the gradual shift of the 4000 \AA\ break from the blue to the red edge of the SDSS \gp\ filter (see Figure \ref{input spectroscopic}). At the lowest redshifts (below z=0.1), the approximately constant photometric redshift uncertainties of $\sigma_{z} \approx 0.01$ imply substantial fractional uncertainties in distance, causing correspondingly large uncertainties in derived parameters. Beyond z = 0.1, these uncertainties in distance fall below 10\%, and decrease with increasing distance. In addition, clusters present at these low redshifts can be more reliably selected using spectroscopic clustering algorithms \citep{mil05}. At redshifts beyond 0.3, the 4000 \AA\ break begins to cross into the \rp\ filter and for a range of redshift from z = 0.32 to z=0.37, there is a significant increase in the photometric redshift uncertainty. 

Focusing on the redshift range from 0.1 to 0.3 has another advantage. The cluster detection methods described here rely on identifying cluster members brighter than 0.4 \lstar\ at the cluster redshift. Our input galaxy catalog is essentially complete for these galaxies out to a redshift of 0.4, where they have \ip\ magnitudes of about 20.5. Cluster galaxies at redshifts less than 0.3 have photometric measurement errors in the essential \gp, \rp, and \ip\ bands which rarely exceed 10\%. This combination allows our cluster selection across this redshift range to be remarkably uniform and allows us to assemble a catalog that is close to volume limited. Residual uncertainty in the evolution of the cluster galaxy luminosity function from z=0.1 to 0.3, which effects our definition of 0.4 \lstar, is one of the principle remaining systematic errors at this point.

\begin{figure}
\rotatebox{90}{\scalebox{0.7}{\plotone{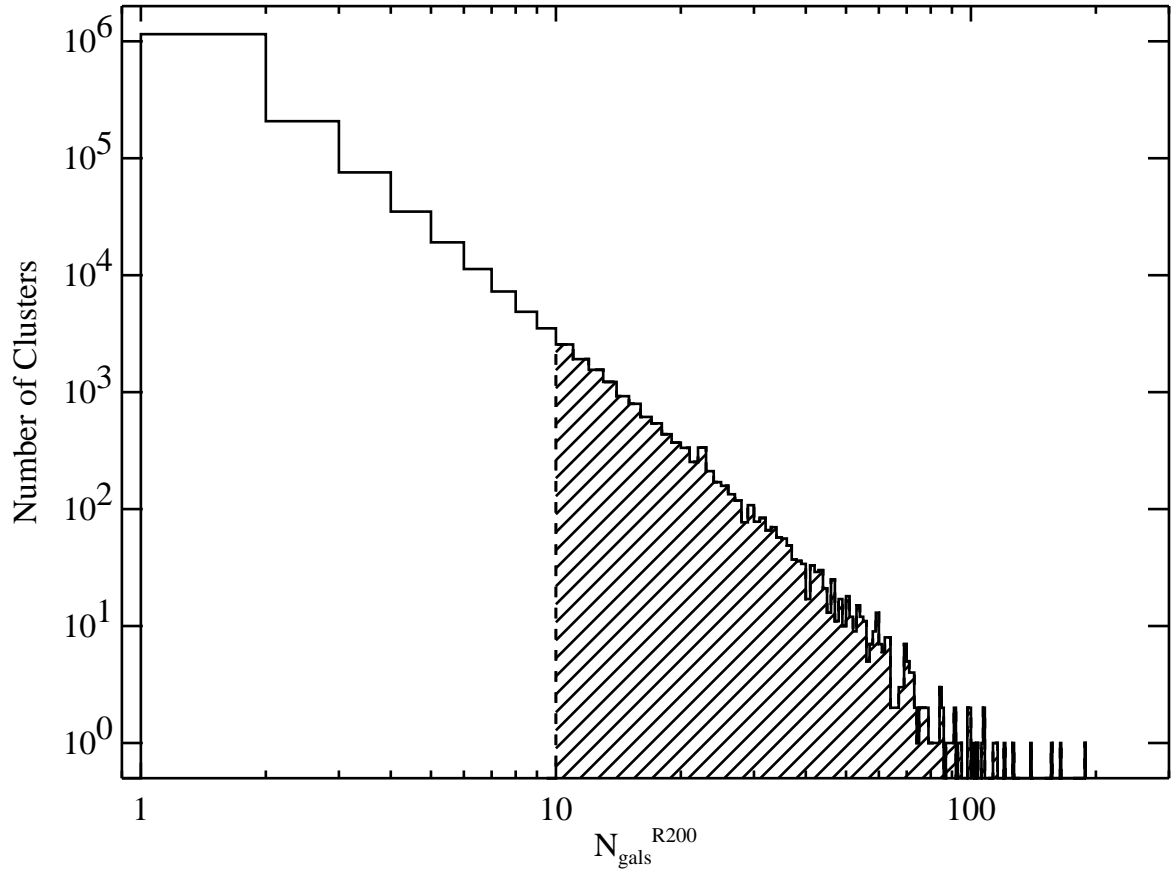}}}
\figcaption{This histogram displays the differential number counts for clusters as a function of richness for the full maxBCG group and cluster catalog (the solid line). The cluster catalog presented here includes all objects in the shaded region. \label{richness distribution}}
\end{figure}

The catalog presented here includes 13,823 clusters with photometric redshifts $0.1 < z < 0.3$ and richnesses of $10 \le $\ngrtwo$ \le 190$; 2891 of which have \ngrtwo\ $\ge 20$. The distribution of these clusters in scaled richness \ngrtwo\ is shown in Figure \ref{richness distribution}. As expected, smaller systems dominate the abundance function. The largest systems are often well-known Abell, Zwicky, or X--ray selected clusters. The \ngrtwo $\ge 10$ objects contain 213,016 member galaxies, or a bit less than 1\% of all input galaxies. Note that the input catalog is apparent magnitude limited, rather than volume limited. So these figures do not transparently reflect the probability that a galaxy will reside in a rich cluster. 

\begin{figure}
\rotatebox{90}{\scalebox{0.7}{\plotone{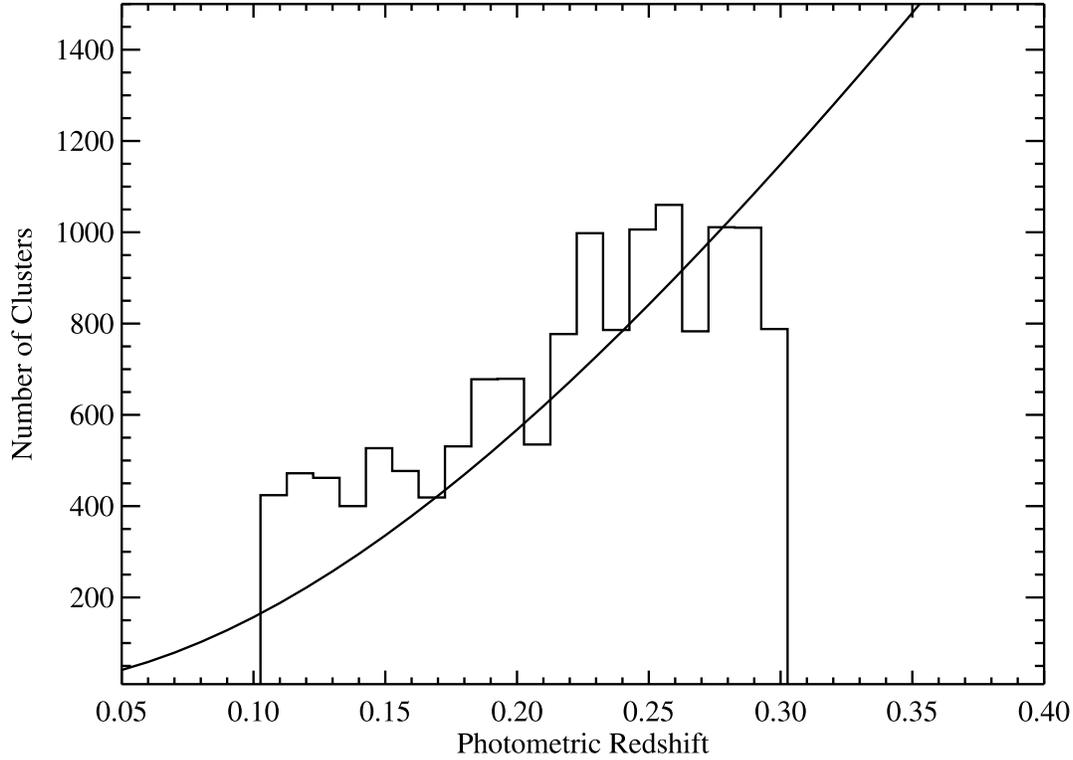}}}
\figcaption{This figure shows the number of clusters as a function of redshift for the maxBCG cluster catalog. The solid line shows the expectation for a volume-limited sample with a density of $2.3 \times 10^{-5}$ clusters $h^{3}$ Mpc$^{-3}$ in a standard $\Lambda$CDM cosmology. \label{redshift distribution}}
\end{figure}

\begin{figure}
\rotatebox{90}{\scalebox{0.7}{\plotone{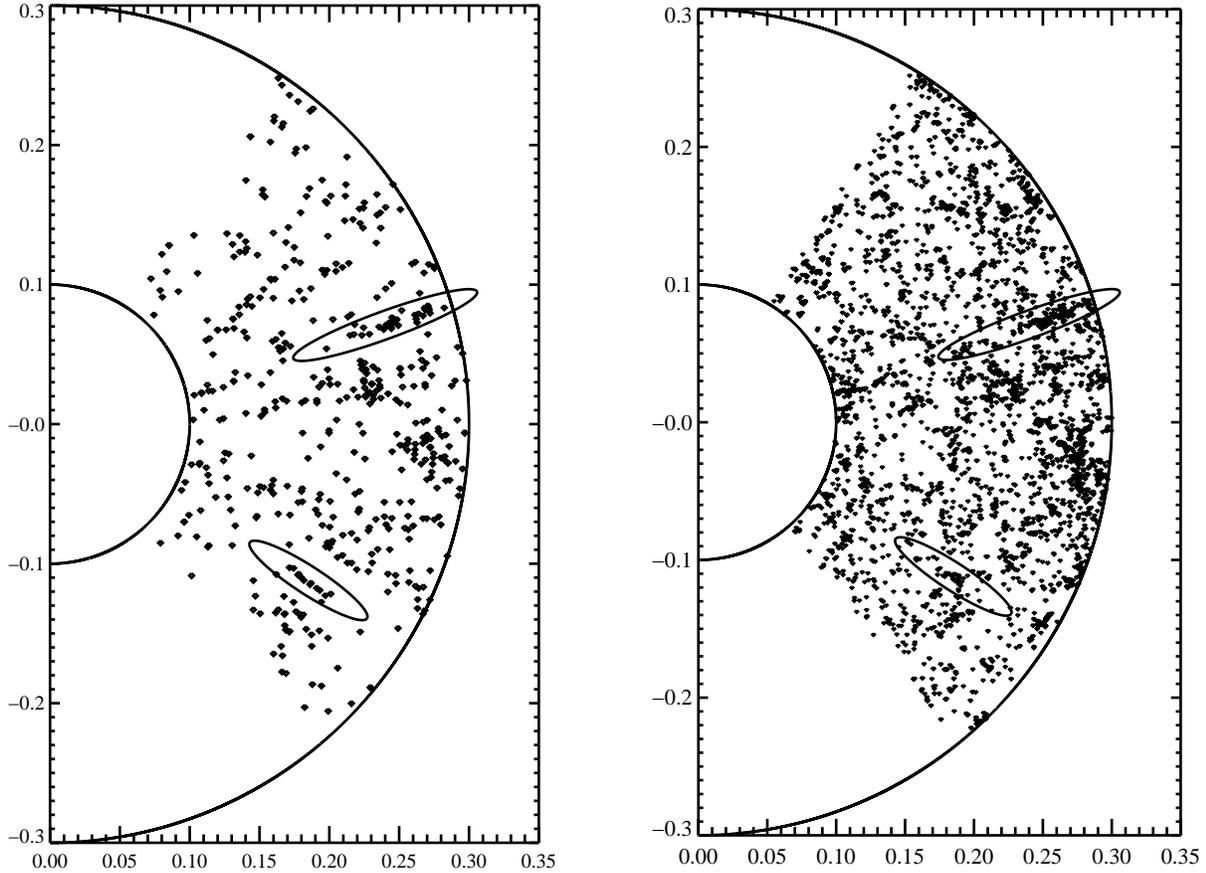}}}
\figcaption{SDSS data in the region -1.25\dsym\ $\le$ Dec $\le$ 1.25\dsym and with RA $<$ 100\dsym\ or RA $>$ 300\dsym, the SDSS southern equatorial stripe, is displayed in this `pie diagram'. The left hand panel shows the locations in RA and photometric redshift of maxBCG cluster centers. The right hand panel shows RA and spectroscopic redshift of SDSS luminous red galaxies (LRGs). This slice contains 492 clusters, about 3.5\% of the total catalog. Circles drawn at z=0.1 and 0.3 show the boundaries of the maxBCG catalog redshift range. The approximately volume-limited nature of the cluster catalog is apparent. There are several features which look like `fingers of god'. These are partly generated by the $\pm$0.01 uncertainties in the photometric redshifts. Their appearance is exaggerated when they enhance real features in the galaxy distribution, as in the two examples outlined by the ellipses added to both figures. \label{pie plot}}
\end{figure}

Among these photometrically selected clusters, a total of 5413 (39\%) have BCGs with measured spectroscopic redshifts. Only a small fraction, mostly at redshifts less than 0.15, have redshifts for more than 5 member galaxies. The lack of multiple redshifts per cluster at high redshifts is primarily due to the magnitude limited nature of the SDSS spectroscopic survey. The photometric redshift distribution of these clusters is shown in Figure \ref{redshift distribution}. Comparison to the expectation for a constant comoving density sample in the standard $\Lambda$CDM cosmology in this figure illustrates the approximately volume-limited nature of the catalog. A `pie diagram' for the clusters located in a 2.5\dsym\ thick slice along the southern celestial equator is shown in Figure \ref{pie plot}, and the volume-limited nature of the sample is again evident.

The full cluster catalog is available as FITS binary and ASCII tables in the online edition of this work. A description of the information provided for each cluster is presented in Table \ref{total_catalog}.

\section{Tests of Cluster Selection in Simulated Data \label{mocks}}

The quality of a cluster catalog is often explored in terms of its false positive rate (purity) and its failure rate (completeness). Until recently, determining completeness and purity has only been possible through first order techniques, for example through random insertion of model clusters into real background data \citep{dia99, ada00, pos02, kim02, got02}. Modern mock galaxy catalogs, in which the galaxy distribution is designed to represent the underlying dark matter distribution from n-body simulations, embed their galaxy clusters in their full environment of filaments and voids, and provide realistic estimates of both the spatial and dynamical structure of the cluster galaxy population \citep{whi02, koc03, eke04, mil05, yan05, ger05, wec06}. These mock catalogs have enabled a deeper exploration of the purity and completeness of catalogs produced by a galaxy cluster finder. Since clusters of galaxies trace the underlying dark matter halo population in the universe, the mock galaxy catalogs can also help to reveal how well the objects found by the cluster-finder relate to dark matter halos. In the end, it is the distribution of these halos in the real universe that we wish to uncover, and the extent to which this distribution is recovered from clusters in the mock catalogs is a strong indicator of the catalog's quality. This recovery rate also provides useful feedback for adjustments and corrections that need to be made to optimize cluster finding algorithms.

To test the cluster finder described in this paper, we utilize the mock galaxy catalogs described by \citet{wec06}, which were largely designed for this purpose. The construction of these catalogs begins with an N-body realization of the large scale structure in the nearby universe. Galaxies are then inserted into this simulation, adopting the locations and motions of dark matter particles, subject to a variety of empirical constraints. The first constraint simply determines the number and nature of the galaxies to be inserted. In this case, the number of galaxies and their distribution in \rp\ luminosity is taken from the measured SDSS galaxy luminosity function \citep{bla01, bla03c}. Galaxies down to 0.4 \lstar\ in \rp\ are then assigned to particular dark matter particles according to a scheme which matches the observed luminosity-dependent two-point clustering of SDSS galaxies as measured by \citet{zeh04}.   The assignment does not
explicitly utilize any knowledge of the locations of dark matter
halos, but produces a halo occupation that is in good agreement with
methods that constrain it directly.  It has the advantage that it
provides a way to naturally include the dim background population necessary for
comparison to photometric data with less resolution than would be
required for a method populating dark matter halos directly.

Once a location and \rp-band luminosity are chosen for each galaxy, appropriate galaxy colors are assigned by selecting a real SDSS galaxy with the same luminosity and measured local density. This process insures that the simulated galaxy colors and their relation to their local galaxy density accurately reflects those found in the data. This is essential if the mock catalogs are to be useful for testing an E/S0 ridgeline based cluster finder. If the dark matter particle on which a galaxy is placed is within \rtwo\ of the center of a dark matter halo, the galaxy is considered a member galaxy of that halo. 

The most serious flaw of this process is that it includes no explicit mechanism for creating the brightest cluster galaxies which we know often rest in the center of a dark matter halo. In a naive effort to fix this, the brightest galaxy in each halo is, as a final step, moved to the center of the halo and assigned the mean halo redshift. It may
be possible to solve this problem by explicitly associating galaxies
with resolved dark matter substructures within halos, as has been done
in recent high-resolution simulations \citep{nag05, con05}, although
it is difficult to do this in simulations with the volume necessary
for photometric cluster studies.

Straight-forward techniques are employed to quantify the purity and completeness. Clusters in the mock catalog are identified using the same code that is run on the data. To understand the false-positive rate (purity), halo membership of galaxies identified as red-sequence cluster members is used to determine the cluster-halo correspondence. In this comparison, it is important to recall that galaxies identified as cluster members by the cluster-finder {\it must} be red-sequence members. As such, they are only a subset of the actual halo members, some of which fall outside the E/S0 ridgeline. This fact affects the matching process, guaranteeing that no perfect one-to-one match of cluster and halo members will be found.

\begin{figure} 
\begin{center}
\rotatebox{0}{\scalebox{0.7}{\plotone{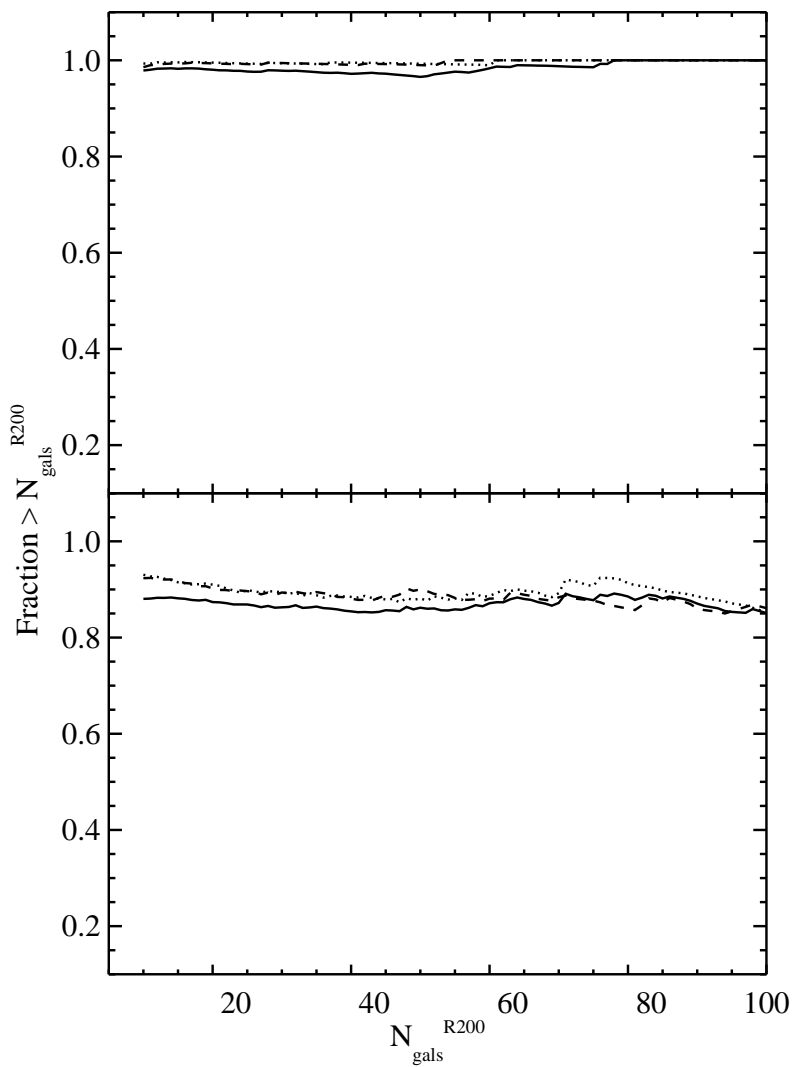}}}
\figcaption{Basic results of purity tests based on mock catalog studies. In each plot the solid line is for the full mock catalog, the dotted lines for halos of 0.1 $<$ z $<$ 0.2 and dashed lines for halos at 0.2 $<$ z $<$ 0.3. The top panel is a purity plot for cluster matching fractions $f_c=0.3$, the bottom for $f_c=0.5$. In each case a cluster is called `real' if a fraction of at least $f_c$ of its E/S0 ridgeline members is contained within \rtwo\ of a single dark matter halo. \label{basic purity}}
\end{center}
\end{figure}

To measure purity, we look to see if each identified cluster corresponds to any real dark matter halo. For each cluster, the halo containing the maximum fraction of a cluster's members is located. If this fraction is less than some threshold, $f_c$, there is no halo to which this cluster clearly corresponds, and the cluster is called a false positive. The exact choice of $f_c$ is arbitrary, and may be driven by the science questions that the catalog is designed to answer. The purity results, presented in Figure \ref{basic purity}, are based are based on two possible choices: $f_c=0.3$ and $0.5$ . At $f_c=0.3$, the catalog is $>90 \%$ pure for $N_{gal}^{r200} > 10$ and $95-100\%$ pure for $N_{gal}^{r200}>20$. Tightening $f_c$ to 0.5 reveals a degraded purity, which is approximately constant at $90\%$ across the full richness range. 

\begin{figure} 
\begin{center}
\rotatebox{0}{\scalebox{0.7}{\plotone{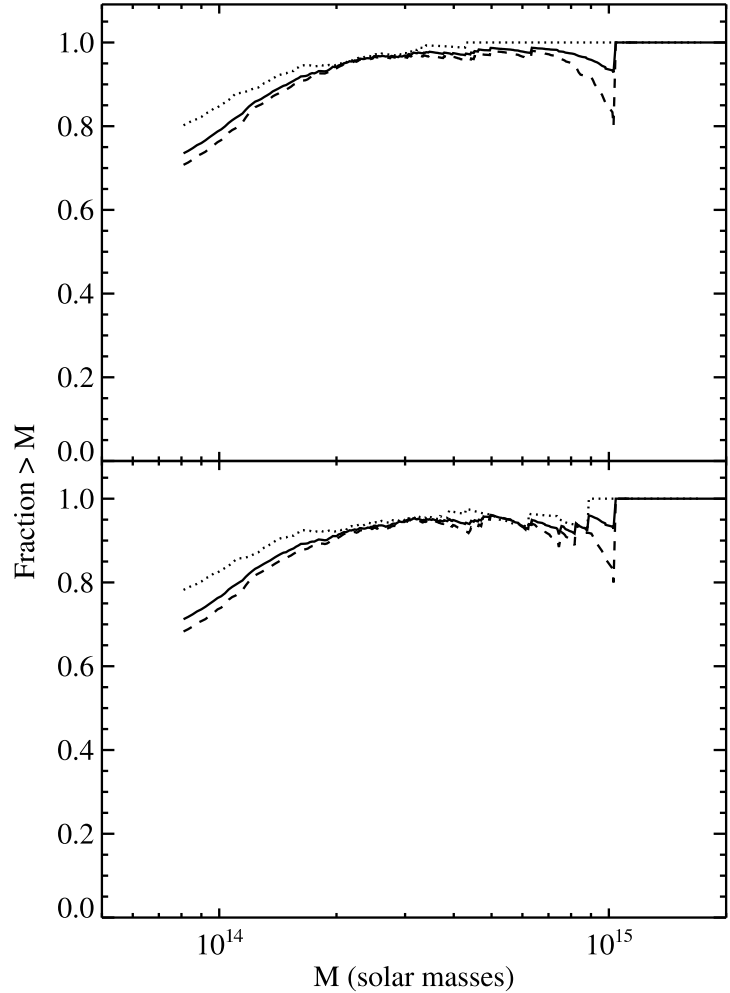}}}
\figcaption{Basic results of completeness tests based on mock catalogs. In each plot the solid line is for the full mock catalog, the dotted lines for halos of 0.1 $<$ z $<$ 0.2 and dashed lines for halos at 0.2 $<$ z $<$ 0.3. The top panel is a completeness plot for a halo matching fraction $f_h=0.3$, the bottom for $f_h=0.5$. In each case, a dark matter halo is considered found if a fraction $f_h$ of its red sequence members is found in a single identified cluster. \label{basic completeness}}
\end{center}
\end{figure}

To evaluate the completeness, the matching process is reversed. For each halo, we identify the cluster that contains the largest fraction of its members. If this fraction is less than some threshold, $f_h$, the halo has been missed. Because the member galaxies in the mock halos are red $and$ blue (and the cluster-finder only identifies the red ones as cluster members), this is a rather strict test. Figure \ref{basic completeness} shows the results of these schemes, in which $f_h=0.3$ and $0.5$, respectively. The sample is $>$90\% complete above $\approx 2 \times 10^{14}$ \hinv \msun\ in the $f_h=0.3$ test, and is 95-100\% complete for objects more massive than $3 \times 10^{14}$ \hinv\ \msun. Notably, both choices of $f_h$ yield nearly 100\% completeness greater than $8 \times 10^{14}$ \hinv\ \msun, with the completeness degrading more rapidly in the larger $f_h$ fraction. In this more stringent test, 90\% completeness is not reached until $\simeq 3 \times 10^{14}$ \hinv\ \msun.

Insight into the cluster-halo correspondence can be gained from considering the meaning of $f_c$ and $f_h$ \citep{ger05}. In measurements of purity, we seek a halo which contains a substantial fraction ($f_c \ge 0.3$) of the cluster's members. When this fraction is small, the cluster-finder has merged one or more additional halos with this best match, a failure we might describe as over-merging. The cluster members beyond those in the matched halo are actually members of other, nearby halos. If a cluster finder were tuned to maximize purity alone, it would endeavor to include as many members of the matching halo as possible, and in the process would be more likely to merge galaxies from neighboring halos. Increasing the value of $f_c$ required for matching ensures that a halo becomes more dominant in the cluster, and that the cluster is not a conglomeration of many smaller halos, until $f_c \simeq 1$. At this point it is possible that the $cluster$ is only part of a much larger $halo$, and there may in fact be several clusters for which this halo is a best match. Similar flexibility in the choice and meaning of matching parameters affects completeness measures. In these, a matching cluster must contain a substantial fraction ($f_h \ge 0.3$) of the halo's total member list. The other halo galaxies may be distributed among other clusters, in which case the cluster-finder has fragmented the halo into several objects. Letting this fraction approach one reduces the incidence of fragmentation, but is likely to cause over-merging. A more extensive discussion of these issues is presented in the companion algorithm paper \citep{koe06}.

Above \ngrtwo = 10 and $M=2 \times 10^{14}$ \hinv\ \msun\, the catalog
presented here is $\simeq 90\%$ pure and complete. At masses below $
\simeq 1 \times 10^{14}$, the decline in completeness is not
surprising. The mean number of real halo members within a three
dimensional \rtwo\ from mock catalogs at masses near $\simeq 1 \times
10^{14}$ is about 8, with a tail to higher richness. When the cluster
finder is run on these halos, it typically increases this true member
number by $\approx$ 20\% due to projection, pushing most $10^{14}$
\msun\ halos over the detection threshold.  Note that the precise
values of purity and completeness are sensitive both to how the
matching is done and to how the limits of the catalog are defined.
When completeness is defined without the explicit threshold, and using
an exclusive matching scheme between halos and clusters, the catalog
is $\simeq 95\%$ complete above $M= 1 \times 10^{14}$ \hinv\ \msun\
(Rozo et al, in preparation).

Halos at lower masses include a wide variety of galaxy groups. Some include clear red-sequences and are easily detected. Others have larger blue fractions and are more difficult to identify by red-sequence techniques. There are indications from the radial profiles of these objects that low \ngrtwo\ objects come from underdense regions of the Universe \citep{han05}, and may be dominated by fossil groups. The mix of objects found by red sequence methods at low \ngrtwo\ makes this an interesting area of future study, as do the objects missed by these means.

An important reason to assemble a cluster catalog is to determine the
distribution of dark matter halos at different masses. Ideally, there
would be a one-to-one correspondence between halos in the real
Universe and observed clusters of galaxies, with the correspondence
encoded in a richness-mass relation such as M(\ngrtwo). Calibrating
this relation would then allow extraction of the true halo
distribution from the cluster abundance function. In fact any
cluster-finder inevitably fragments some halos into sets of smaller
clusters, and merges some smaller halos into larger clusters. This
complicates cluster-to-halo matching, and illustrates the important
role which studies of realistic mock catalogs must play: without such
studies, reliable cosmological constraints from optically-selected
catalogs are likely impossible. A much more complete analyses of
fragmentation, merging, and the richness-mass relations required to
generate final estimates of the halo mass function will be presented
in Rozo et al (in preparation).

\section{Tests of Derived Cluster Parameters \label{tests}}

In this section we describe some tests of the quality of derived cluster properties, based in this case primarily on reference to SDSS spectroscopic data. The large spectroscopic catalog provided by the SDSS (Figure \ref{input spectroscopic}) enables studies of the photometric redshift quality, the effects of projection on the cluster finder, and of the richness-velocity dispersion relation. 

\subsection{Tests of Photometric Redshifts}

\begin{figure}
\rotatebox{90}{\scalebox{0.7}{\plotone{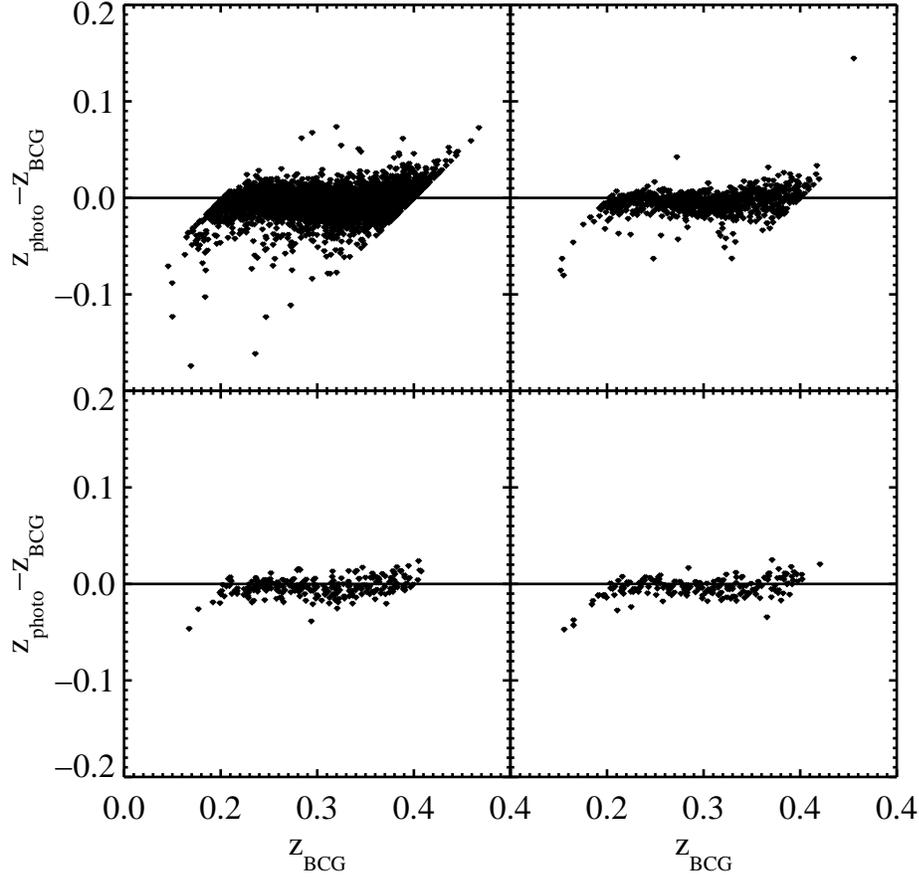}}}
\figcaption{The comparison of BCG spectroscopic redshift, $z_{BCG}$, to cluster photometric redshift, $z_{photo}$, for 5413 objects, divided into several different richness \ngrtwo\ ranges: $10 < N_{gals}^{r200} < 20$ (upper left), $20 < N_{gals}^{r200} < 30$ (upper right), $30 < N_{gals}^{r200} < 40$ (lower left), $40 < N_{gals}^{r200}$ (lower right). Note the one object in the upper right panel which has a spectroscopic redshift $\approx$0.35 and a photometric redshift $\approx$0.21. This is the BCG of a cluster actually at z=0.21. The spectrum of the central galaxy (SDSS J075137.2+325447.4) shows a typical red galaxy spectrum with z=0.21 overlaid with strong emission lines from a (possibly lensed) background object at z=0.355. In this case, the photometric redshift in the cluster catalog is correct. \label{photoz scatterplot}}
\end{figure}

\begin{figure}
\begin{center}
\rotatebox{0}{\scalebox{0.7}{\plotone{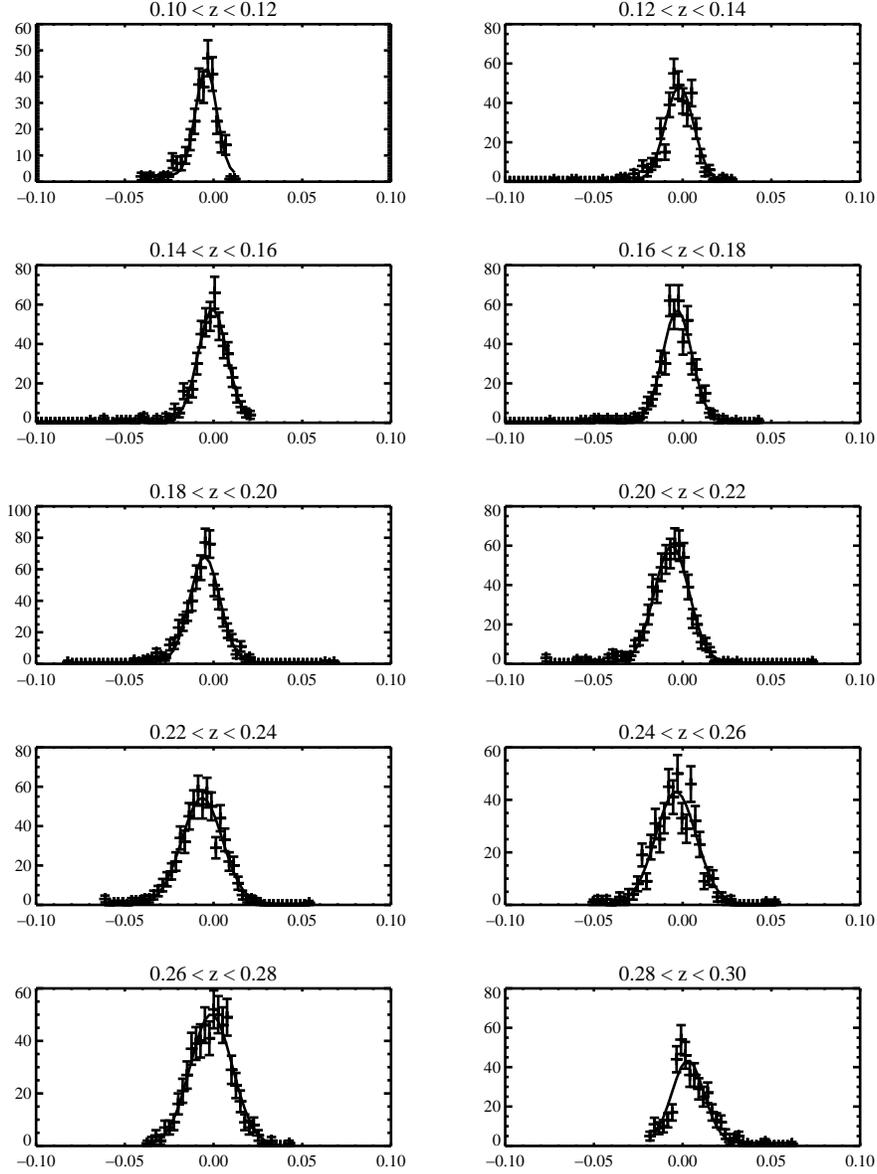}}}
\figcaption{The difference between measured BCG spectroscopic redshift and cluster estimated photometric redshift for a total of 5413 clusters with \ngrtwo\ $\geq$ 10 for which spectroscopic BCG redshifts exist. Each panel shows the difference between spectroscopic and photometric redshift for a small bin of spectroscopic redshift. Gaussian fits to each distribution are overlaid. The fit dispersions range from $\sigma_z$=0.006 to $\sigma_z$=0.011. The small ($\Delta_z \approx 0.004$) average bias seen here is subtracted in the final catalog. \label{bcg photoz}}
\end{center}
\end{figure}

Determining the physical properties of observed clusters requires accurate photometric redshifts. To assess the quality of $z_{photo}$ for a particular cluster, its true redshift $z_{spectro}$ must be known from spectroscopy. For this test of photometric redshifts, we use the clusters with BCG spectroscopic redshifts. Since the BCGs are bright, they are more likely to be targeted for spectroscopy than an average cluster galaxy. In addition, the SDSS LRG target selection \citep{eis01} explicitly constructs a volume-limited sample of the brightest red galaxies, many of which are BCGs. The combination provides good statistics for the evaluation of $z_{photo}$. Figure \ref{photoz scatterplot} compares all available BCG spectroscopic redshifts ($z_{BCG}$) to cluster photometric redshifts ($z_{photo}$) for objects in four different \ngrtwo\ richness ranges. Figure \ref{bcg photoz} shows the comparison of $z_{photo}$ to $z_{spectro}$ in a series of different redshift ranges. The dispersion in $z_{spectro} - z_{photo}$ is $\simeq 0.01$, essentially independent of redshift. There is a small bias ($\approx 0.004$), in the sense that estimated redshifts are slightly higher than measured ones at all redshifts. We correct for this small bias in the photometric redshift estimates provided in the final catalog. The errors throughout are well represented by a Gaussian. Photometric redshifts for the richest half of the clusters are slightly better, with a typical dispersion in $z_{spectro} - z_{photo}$ of about $0.008$. This photometric redshift accuracy allows clusters separated by more than about 50 \hinv\ Mpc along the line of sight to be correctly identified.

\subsection{Tests of Projection}

The projection of foreground and background galaxies onto the observed cluster population is a concern for all optical cluster selection algorithms. There are two levels at which projection occurs. First, there is large scale projection, in which galaxies more than a few tens of Mpc from a cluster and randomly projected on it are mistaken for cluster members. This large scale projection can easily be tested for using spectroscopic redshifts, and is an effect one might aspire to completely avoid. Second, there is small scale projection, in which galaxies near the cluster, along infalling filaments for example, are thought to lie within the cluster. Spectroscopic information about cluster membership on these small scales is limited by redshift space distortions. This kind of projection is best constrained using realistic simulations. 

By relying on the tight clustering of E/S0 galaxies in color--magnitude space, the maxBCG cluster finder limits the influence of projection, but it remains important to quantify this source of contamination. Direct study of projection also provides insight into the limitations of red-sequence cluster-finding and may suggest algorithmic improvements. 

To study the incidence of projection in this catalog, we take advantage of the large number of spectroscopic redshifts available for both BCGs and member galaxies identified by the algorithm. We begin by identifying the 3057 clusters which have spectroscopic redshifts for both the BCG and at least one member galaxy. The first step is to define a ``best'' redshift for each of these clusters, ``$z_{best}$''. When a cluster has at least 3 member galaxy spectra, we take the median redshift of all the cluster member galaxies as $z_{best}$. If there are fewer than 3 member redshifts, we define $z_{best}$ to be the BCG redshift. To test for projection we then compare individual BCG and member spectroscopic redshifts to these estimates of $z_{best}$. Our determination of $z_{best}$ is far from perfect; often it is measured from just a few member galaxy redshifts or from only the redshift of the BCG. Because of this, we further split the spectroscopic sample into the 143 clusters with at least 10 member galaxies located within $\pm$ 2000 km~s$^{-1}$ of the median redshift and the remaining 2914 clusters with fewer. Comparing the projection results obtained with these two samples provides a check on the influence of uncertainty in the determination of $z_{best}$ on our conclusions.

\begin{figure} 
\rotatebox{90}{\scalebox{0.7}{\plotone{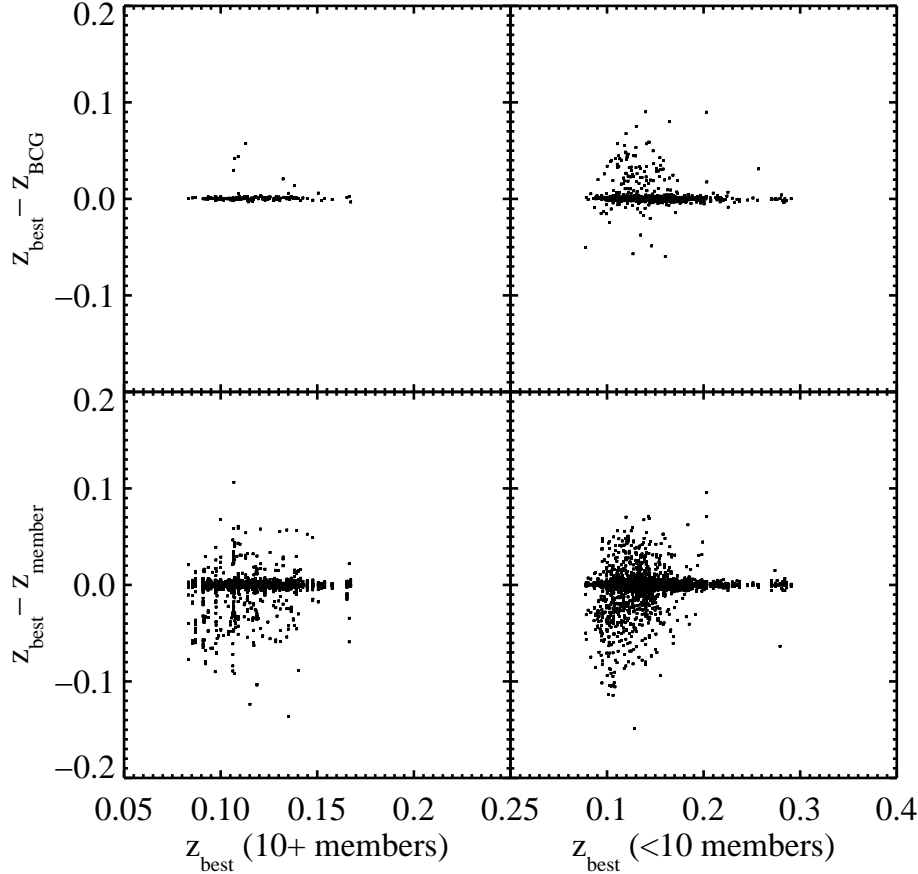}}}
\figcaption{Residual plots of BCG ($z_{BCG}$) and member ($z_{member}$) projection in maxBCG clusters. The top left panel compares BCG redshifts to median member redshifts, $z_{best}$, in the 143 cases where at least 10 member galaxies are found within $\pm$ 2000 km~s$^{-1}$ of the median. Only 8/143 (5.6\%) of these BCGs are found with velocities differing from the median by $>$ 2000 km~s$^{-1}$. The upper right panel is the same comparison for the remaining 2914 clusters, for which the apparent BCG projection fraction is 117/910 (13\%) (excluding those where $z_{best}=z_{BCG}$). The lower panels show similar comparisons for all member galaxies with spectroscopic redshifts. The lower left compares $z_{member}$ to $z_{best}$ for the 143 well-measured clusters, while the lower right does the same for the remaining clusters. A fraction 427/2701 (16\%) of members are seen to be projected in the well-measured cases, and 829/5068 (16\%) in the remainder. \label{projection tests}}
\end{figure}

We first measure the incidence of BCG projection. Since we require the BCG to be the brightest galaxy in a cluster, one might worry that we would occasionally select as a BCG a bright foreground galaxy projected onto the E/S0 ridgeline of a more distant cluster. To test for projection, we compare $z_{BCG}$ to $z_{best}$ in the top two panels of Figure \ref{projection tests}. The top left panel shows the comparison for the 143 clusters with the best determined median member redshifts, while the top right panel shows the comparison for the remainder, including only those with between 3 and 10 member galaxies with spectra; those with $z_{best}=z_{BCG}$ are removed altogether. The fraction of BCGs with velocities differing from the median by more than 2000 km~s$^{-1}$ is 5.6\% (8/143) for the best-measured set and 13\% (117/910) for the remainder. While BCG projection does occasionally occur, it is not particularly common. When it does happen, it is most common to find an unusually red foreground galaxy projected onto a more distant cluster. BCG projection is also somewhat less likely as clusters increase in richness.

To study the importance of projection for cluster member galaxies, we compare individual member redshifts, $z_{member}$, to the best estimate of the cluster redshift, $z_{best}$. Again, we conduct the comparison both for the clusters with the best determined $z_{best}$ and also for the remainder, including only those with between 3 and 10 member galaxies with spectra.. These comparisons are shown in the lower two panels of Figure \ref{projection tests}. In the best-measured clusters, 16\% (427/2701) of member galaxies are found with velocities differing from the median by more than 2000 km~s$^{-1}$. For the less well-measured clusters, 16\% (829/5068) have velocities differing from the median by more than 2000 km~s$^{-1}$. Since the determination of $z_{best}$ is not especially robust for these clusters, this represents an upper bound on the degree of member projection.

This test demonstrates the relative insensitivity of red-sequence methods to large-scale projection effects. No more than 13\% of BCGs and $16\%$ of the galaxies identified as cluster members according to color and spatial location are actually observed in projection. The majority of projected members lie behind the cluster; they are intrinsically bluer galaxies at higher redshift. Narrowing the color window for membership can reduce the incidence of projection, but as shown in the accompanying algorithm paper \citep{koe06}, it also increases the fragmentation of dark matter halos. 

The conclusions drawn here are restricted somewhat by the realities of the SDSS spectroscopic selection. Galaxies for which spectra are obtained are preferentially bright cluster members, and there is a small fiber-collision bias against the measurement of galaxies in the densest regions. Since the member galaxies selected here are all brighter than 0.4 \lstar\ and the probability of cluster membership is highest in the densest regions, the effect of these limitations on our conclusions should be minimal, but complicates measurements such as the radial dependence of member projection \citep{mck06}.

\subsection{Relating Richness and Velocity Dispersion}

In addition to their usefulness in confirming cluster redshifts, spectroscopic samples of galaxies can be applied to understand the relation between richness and cluster velocity dispersion. In the current catalog, the number of E/S0 ridgeline galaxies, \ngrtwo\, is a basic indicator of cluster richness. To understand the relationship between this cluster richness and velocity dispersion, we begin with a list of clusters for which a spectroscopic redshift for the BCG is known. Around each such BCG, we search for any other galaxies with measured spectroscopic redshifts, and define spectroscopic ``pairs''; each of which includes a BCG and a nearby tracer of the velocity field. For this study we keep all pairs in which the neighboring galaxy lies within a projected distance \rtwo\ of the BCG and within $\pm$7000 km~s$^{-1}$ of the BCG velocity, irrespective of whether the neighboring galaxy is a member of the E/S0 ridgeline; this $R_{200}$ is the same derived from \citet{han05} for photometric data. In many cases, a cluster will contribute only a few pairs, making the determination of individual velocity dispersions impractical. As a result, we gather together all pairs for clusters of similar richness, and hence measure the average pair-wise velocity difference (PVD) structure around a class of similar clusters. This \emph{stacking} approach is analogous to that applied in the study of halo masses in isolated galaxies by \citet{mck02} and \citet{pra03}. 

\begin{figure}
\rotatebox{0}{\scalebox{0.9}{\plotone{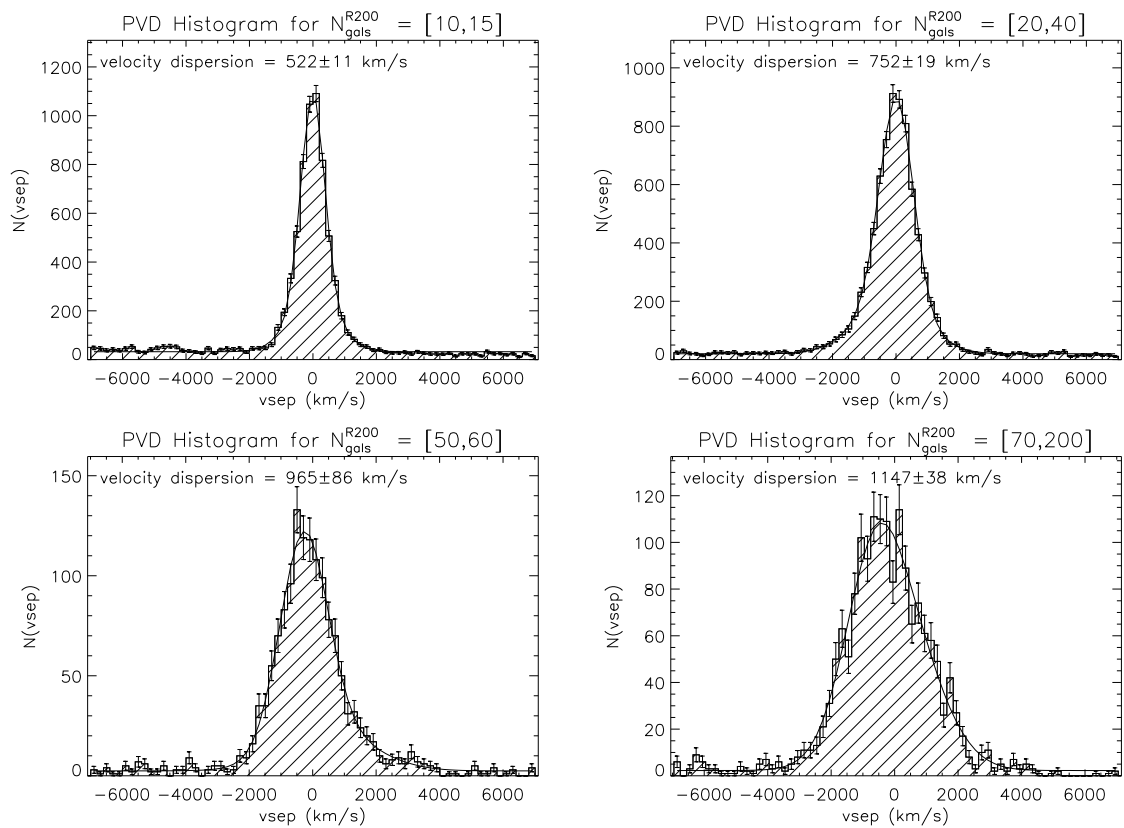}}}
\figcaption{This figure shows examples of the velocity distribution seen relative the the BCGs in clusters with various $N_{gals}^{r200}$. While the distribution is approximately Gaussian, it is much better fit by a combination of two Gaussians. The characteristic dispersions shown in the legends are the appropriately-weighted averages of the two best fit Gaussians. \label{sigma_v examples}}
\end{figure}

Examples of the velocity structure seen around these clusters in several \ngrtwo\ bins are shown in Figure \ref{sigma_v examples}. These PVD histograms are approximately Gaussian, but show clear evidence for components both narrower and broader than the average. This probably arises from the imperfect mapping between \ngrtwo\ and mass. If all clusters in a narrow bin of \ngrtwo\ had the same mass, we would expect a very nearly Gaussian PVD. If, however, the bin contains a mix of both lower and higher mass objects, the PVD will include a mix of narrower and broader components \citep{sco04}. To determine the typical velocity dispersions in each bin, we fit two Gaussians plus a constant to the PVD histogram. The constant is shown by \citet{woj02} to adequately account for unbound particles in simulations \citep[see also][]{mck06}. The dispersion is then square root of the weighted average of the variances of the two Gaussians.  The weights of each Gaussian are determined in the fitting procedure (the Expectation Maximization algorithm, see \citealt{con00}). Two Gaussians are used in the fit more as a convenient and stable parametric fit to the PVD histogram. They allow us to fully measure the second moment of the PVD histograms and more importantly, to measure their kurtosis.  The kurtosis is an important parameter that allows one to constrain mass mixing in these PVD histograms.

\begin{figure}
\rotatebox{0}{\scalebox{0.9}{\plotone{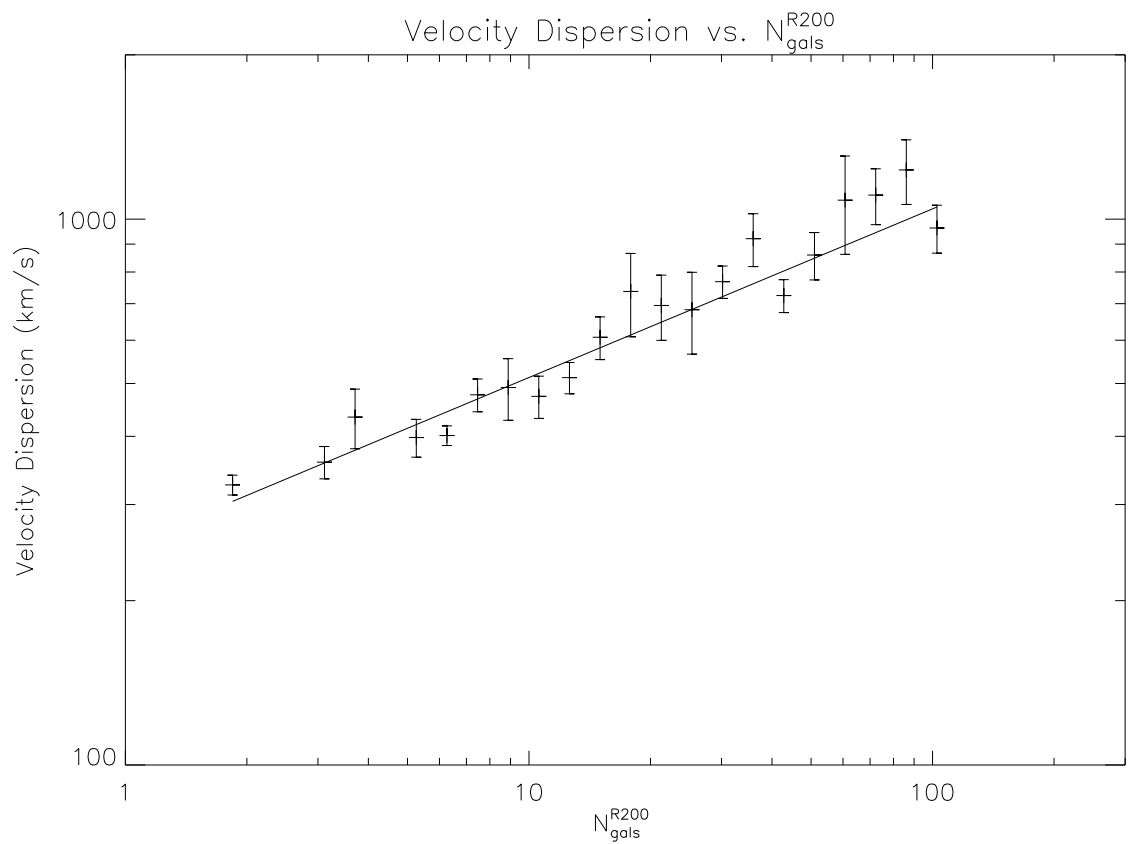}}}
\figcaption{This figure shows the dependence of stacked velocity dispersion on scaled richness $N_{gals}^{r200}$. The steady increase of stacked velocity dispersion with richness illustrates the connection between $N_{gals}^{r200}$\ and mass. The best fit power law for this data is $\ln{\sigma} = (5.52\pm0.04) + (0.31\pm0.01) \ln{N_{gals}^{R200}}$. At the $N_{gals}^{r200}$ $\geq$ 10 threshold for this cluster catalog, the typical velocity dispersion is $\approx 500$ km~s$^{-1}$. \label{sigma_v v ngrtwo}}
\end{figure}

Median velocity dispersions for clusters binned by \ngrtwo\ are shown in Figure \ref{sigma_v v ngrtwo} . There is a clear increase in measured velocity dispersion with richness \ngrtwo. This increasing velocity dispersion is related to the increasing mass of the sample, reinforcing the idea that \ngrtwo\ is a useful proxy for mass. This richness measure is simple; just a count of the number of bright E/S0 ridgeline galaxies in a cluster. A variety of refined richness measures can be derived from this data, including matched filter likelihoods, member galaxy luminosities, and total optical luminosities. It is likely that combinations of luminosity and concentration measures will provide useful mass estimators for clusters, just as they do for individual elliptical galaxies \citep{zar06}. It should also be useful to include measures of the diffuse optical light in the clusters, as this may include as much as 40\% of the total optical light \citep{zib05, gon05, kri06}. A more detailed description of these richness-dispersion scaling relations along with their extension to cluster mass and mass scatter estimates is given in a companion paper \citep{mck06}. Additional information about the masses of clusters detected in the SDSS can be provided by weak lensing measurements, as was done for a much smaller cluster sample in \citet{she01}. The combination of lensing and dynamical mass calibration will provide the input needed for estimation of the cluster mass function.

\subsection{Measurements of Space Density}
 
\begin{figure}
\rotatebox{90}{\scalebox{0.7}{\plotone{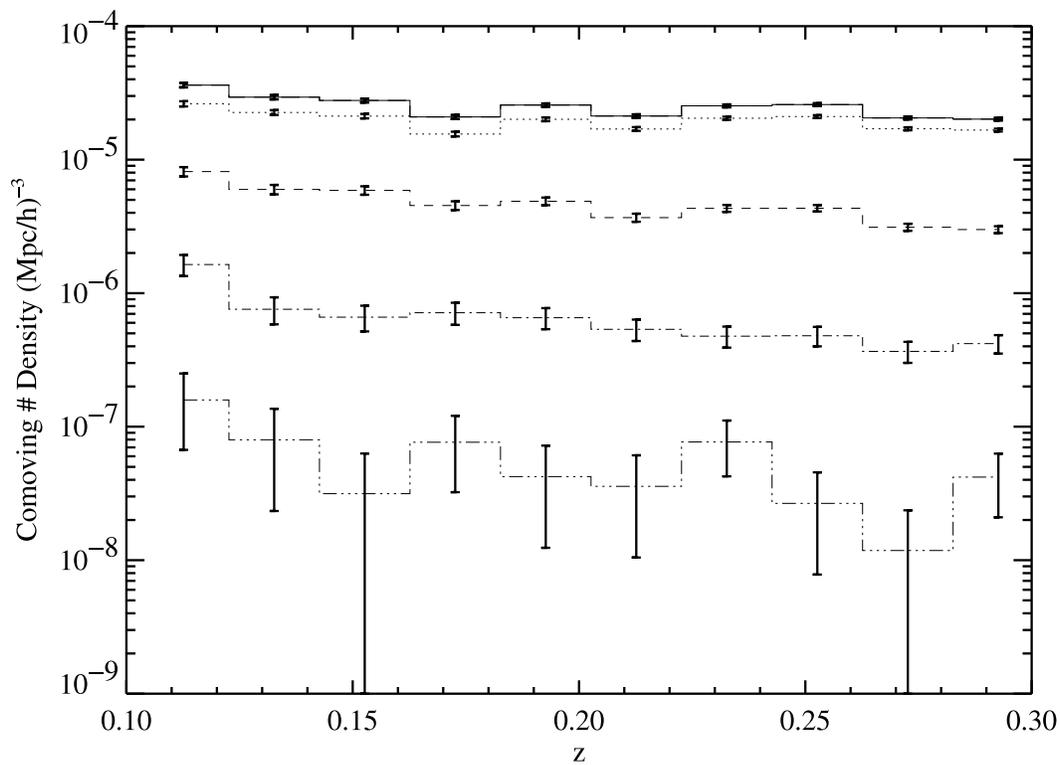}}}
\figcaption{Comoving number density of objects of varying richness. The solid top line is the space density for the full catalog. The lines below represent the space density in ranges of richness. From top to bottom these ranges are: $10 < $$N_{gals}^{r200}$$ < 20$, $20 < $$N_{gals}^{r200}$$ < 43$, $43 < $$N_{gals}^{r200}$$ < 91$, $91 < $$N_{gals}^{r200}$$ < 189$. Poission error bars are overplotted. \label{space density}}
\end{figure}

 $\Lambda$CDM
simulations reveal that the space density of dark matter halos is
nearly constant out to $z \sim 0.3$, with only a small increase in the
number of objects at the high end of the mass function at late
times. In Figure \ref{space density}, the comoving number density of
clusters in a given $N_{gals}^{R200}$ bin is plotted as a function of
redshift. No correction for purity or completeness is applied. 
It is approximately flat in all bins, as predicted by
theory. This is a simple demonstration of the roughly volume-limited
nature of the catalog. Measurements on simulations indicate that the
slight decrease seen in the highest  $N_{gals}^{R200}$ bin is due to evolution
in the mass--$N_{gals}^{R200}$ relation or a redshfit-dependent bias in the definition of $0.4L_*$ and not to incompleteness
at high redshift. Before proceeding to use this measurement as a
tool to constrain cosmology, careful calibration must be made of the
richness-mass relation and especially any possible dependence of this
relation on redshift. The details of this mass calibration will be
presented in future papers on dynamical and weak lensing mass
estimators.

\section{Comparison to Known X--ray Selected Clusters \label{X--ray comparison}}

An important test of any cluster finder is the extent to which it identifies those clusters found by other means. Optically-selected catalogs generated by alternative codes with measurements, sky, and redshift coverage similar to the current catalog are not yet available, but will in the future be invaluable in understanding selection biases in maxBCG. X--ray surveys are attractive for this purpose, in that they are dependent on different cluster physics, have mass proxies, accurate redshifts, and large sky coverage. Just as in comparisons of optically-identified clusters to mock catalog halos, some care must be taken. Optical and X--ray methods may identify different cluster centers, and while optical richness and X--ray luminosity are clearly coupled \citep{lin04, pop05}, there is significant scatter in this relation. Nevertheless, we expect that in the field of X--ray bright clusters there should be some significant overdensity of galaxies that is approximately centered on the X--ray peak. With this in mind, we investigate the extent to which known X--ray clusters were identified in this catalog. Because the maxBCG cluster-finder centers clusters according to the BCG location, we will compare our BCG location to the X--ray peak location. 

Comparison of optical and X--ray catalogs is complicated by several factors. First, central galaxies in clusters are also not always perfectly coincident with peaks in X--ray emission, and the identification of a single cluster BCG in merging systems is often ambiguous. The X--ray brightest clusters are also especially likely to contain BCGs exhibiting unusual colors, often due to cooling flows with accompanying star formation or AGN activity \citep{cra04}. In addition, the optically-selected sample presented here extends to much lower mass objects than existing large area X--ray catalogs. For these reasons, we do not use the X--ray sample to quantify the purity of the optical sample, but merely as a check of our ability to locate most previously identified clusters. 

We conduct our comparison to the combined NORAS \citep{boe00} and REFLEX \citep{boe04} catalogs. These catalogs are purely X--ray selected, flux limited to about $3 \times 10^{-12}$ ergs cm$^{-2}$ s$^{-1}$ at energies $0.1-2.4$ keV, and when combined cover both the Northern and Southern Galactic caps. Within the redshift limits of this maxBCG catalog ($0.1 < z < 0.3$) and its sky coverage, the combined NORAS and REFLEX catalogs contain 99 X--ray clusters. It is important to note that the NORAS catalog was based on early selection of extended sources and as a result is not as complete or pure as the later REFLEX catalog. A new generation of the NORAS catalog is under construction \citep{boe06}.

To determine if an X--ray cluster was ``found'' in our maxBCG catalog, a cylinder centered on the X--ray peak with a 2 \hinv Mpc radius and a depth $|z_{xray}-z_{photo}| < 0.05$ is searched for optical clusters.  For 94 of the 99 X--ray clusters a single, quite rich object and occasionally a few lower richness objects are found within these boundaries. The best match is defined by combining the richness of the optical object and the proximity to the center of the X--ray cluster. Despite the fact that a relatively deep redshift box is used for this match, the matches identified in this way have redshifts which agree with those given in NORAS and REFLEX with a dispersion of $\sigma_z \approx 0.003$, so they are clearly physically related. Because this exercise is automated, it differs from that in some previous comparisons of X--ray and optical properties. Here we compare optical centers identified with an automatic algorithm with X--ray centers, rather than visually examining each cluster and manually identifying BCGs. It shows that in many cases, the center selected by the automatic algorithm agrees with the BCG which would be manually selected. For 76 of these 94 matches, the offsets between X--ray and BCG centers are less than 250 \hinv\ kpc. For these objects, the distribution of offsets has a median of 58 \hinv\ kpc and a standard deviation of 57 \hinv\ kpc. An additional 15 matches have offsets between 250 \hinv\ kpc and 1 \hinv\ Mpc. The remaining three matches have very large separations, between one and two Mpc. The separations larger than 250 \hinv\ kpc occur for a variety of reasons, and each of these cases, along with the five not matched, is described individually below.

\begin{figure}
\rotatebox{90}{\scalebox{0.7}{\plotone{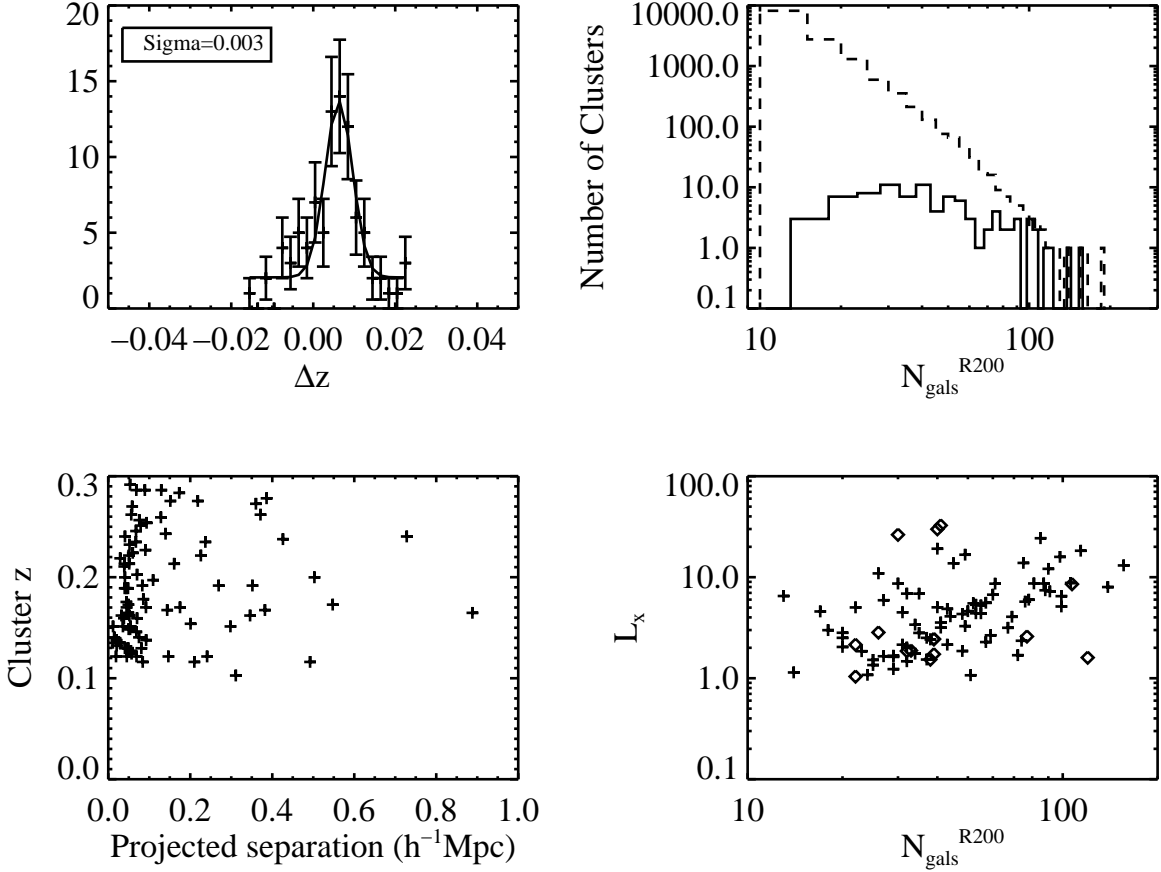}}}
\figcaption{This figure summarizes the results of comparisons of the maxBCG catalog to the combined NORAS and REFLEX catalogs. The upper left shows the difference between NORAS/REFLEX spectroscopic cluster redshifts and the matching maxBCG photometric redshifts. The dispersion of the best fit Gaussian (shown overlaid) is $\sigma_z$=0.004. The upper right shows the richness distribution of the full maxBCG catalog compared to the richness distribution for the clusters which match NORAS/REFLEX sources. The lower left shows the projected separations between X--ray and optical centers as a function of redshift. The lower right compares X--ray luminosity (in units of $10^{43}$ from NORAS/REFLEX to the $N_{gals}^{r200}$\ optical richness. Plus symbols are the 73 clusters with X--ray optical offsets $\le$ 250 \hinv\ kpc, diamonds are the 14 with offsets from 250 to 1000 \hinv\ kpc. \label{xray matching}}
\end{figure}

The X--ray matching is summarized in Figure \ref{xray matching}. This upper left panel in this figure shows the close correspondence between X--ray redshift and maxBCG photometric redshift. The upper right panel compares the overall richness distribution of the maxBCG catalog to the richness distribution of the X--ray matched clusters. While it is clear they are drawn preferentially from high richness objects, the effect is diluted by the flux-limited nature of the X--ray catalogs. The lower left panel compares the projected separation between X--ray and BCG centers to the cluster redshift. The tightly clustered group of matches with separations $\le$ 250 \hinv\ kpc is clear, along with the smaller number of large separation matches. The lower right panel compares the optical richness \ngrtwo\ to the X--ray luminosity for these clusters, drawing a distinction between the close matches and those which are more distant. The correlation is poor, and will be undertaken in a future study. Point source contamination and the chosen physical scale are among the complicating factors.

A case by case comparison was undertaken for each match separated by 250-2000 \hinv kpc. The full details of this comparison are reserved for a future work which examines the relationship between optical and X--ray properties of clusters. The 23 clusters for which the maxBCG cluster location differs from the NORAS/REFLEX X--ray center by 250-2000 \hinv kpc fall into several categories: complications in centering the X--ray flux due to point source contamination and mergers, maxBCG centering complications due to the lack of a dominant BCG, BCGs with evidence of cooling flows which produce rather blue colors, and matched maxBCG clusters with richnesses below the $N_{gals}^{r200}$ threshold. In these cases, the X--ray cluster is not missed; the centers chosen by optical and X--ray means disagree.

A final concern in this matching exercise is the incidence of matches by chance. The large radius matches (250-2000 \hinv kpc) of maxBCG clusters to X-ray clusters may be the most subject to chance projections. To investigate this, we place 13,823 random points in the SDSS footprint with redshifts drawn from the maxBCG redshift distribution and positions non-coincident with maxBCG clusters within a projected $2 h^{-1}$Mpc. The same matching routine is run to match these points to the NORAS/REFLEX X--ray catalog. Of the 99 NORAS/REFLEX X--ray clusters, we find 0 matches between $250 h^{-1}$ and $1000 h^{-1}$kpc, and 6 between $1000 h^{-1}$ and $2000 h^{-1}$kpc. In the same ranges for the maxBCG clusters, we find 18 between $250 h^{-1}$ and $1000 h^{-1}$kpc and 5 between $1000 h^{-1}$ and $2000 h^{-1}$kpc. This simple test indicates that outside $1 h^{-1}$Mpc, our matches are more subject to chance projections with the NORAS/REFLEX X--ray sample.  

This simple matching exercise shows that in the catalog presented here, we are $\simeq 80\%$ successful at automatically identifying the BCGs selected manually by visual selection in bright X--ray clusters. This comparison also highlights some of the difficulties encountered in optical X--ray comparisons. Many X--ray bright clusters have emission line BCGs (perhaps 25\%; \citep{cra99}) at their centers. Cluster-finders that search for red-sequence BCGs may miss these centers. Future cluster finding algorithms based on the red sequence may
choose to account for this by searching for BCGs with unusually blue colors, although doing so invites contamination by foreground galaxies. In addition, there are examples in the X--ray catalogs of clusters without a single unambiguous BCG, and of clusters where the visually apparent BCG does not coincide well with the peak of X--ray emission. In some cases this may be due to contamination of the X--ray signal by cluster galaxy AGN emission. Unrelaxed clusters or those undergoing mergers are especially prone to this. The cases in which the optical cluster finder did not automatically identify the visually selected BCGs come from a mix of these issues. They are a combination of the algorithm used for cluster finding and the complex cluster physics that determine the BCG location, spectrum, the optical galaxy membership, and the X--ray signal.

\section{Discussion \label{discussion}}
We have presented a new catalog of clusters of galaxies selected from SDSS photometric
data using the maxBCG technique, which represents the largest galaxy cluster
catalog assembed to date.  This technique utilizes the
clustering of galaxies on the sky, in magnitude, and color to identify
groups and clusters of bright E/S0 ridgeline galaxies. It identifies
objects ranging in size from single isolated ellipticals to the
richest galaxy clusters. We have extracted from this very large list a
sample of 13,823 clusters containing at least 10 E/S0 ridgeline
galaxies brighter than 0.4 \lstar. Running this cluster finder on
realistic mock galaxy catalogs allows us to show that this cluster
sample is more than 90\% pure and more than 90\% complete for halos
with masses $\geq 2 \times 10^{14} h^{-1} M_{\sun}$. Comparison between
maxBCG photometric redshifts and SDSS spectroscopic redshifts for BCGs
demonstrates the precision and accuracy of the derived cluster
redshifts. The large volume of SDSS spectroscopic data also allows, in
a somewhat limited way, an examinination of the effect of projection
on both our identification of brightest cluster galaxies and on our
determination of cluster membership. In both cases we show that large
scale projection plays only a small role in red-sequence cluster
detection.

We have shown that the basic richness measure presented here, the scaled number of bright E/S0 ridgeline galaxies \ngrtwo\, is strongly correlated with cluster velocity dispersion. Comparison of this optically-selected catalog with the existing NORAS and REFLEX X--ray selected cluster catalogs reveals that nearly all of these relatively X--ray luminous objects are detected among the richest of the optical clusters found here. As comparable wide-angle imaging cluster catalogs with accurate redshifts become available, these comparisons will be extended to catalogs selected by alternate optical algorithms as well.

Further aspects of this cluster catalog, including galaxy populations and profiles, mass calibration by both dynamical and lensing measurements, diffuse light measurements, and the cosmological constraints it can place will be examined in future papers.

\acknowledgments

Funding for the SDSS and SDSS-II has been provided by the Alfred
P. Sloan Foundation, the Participating Institutions, the National
Science Foundation, the U.S. Department of Energy, the National
Aeronautics and Space Administration, the Japanese Monbukagakusho, the
Max Planck Society, and the Higher Education Funding Council for
England. The SDSS Web Site is http://www.sdss.org/. T. McKay,
A. Evrard, and B. Koester gratefully acknowledge support from NSF
grant AST 044327. RHW is supported by NASA through Hubble Fellowship grant
HST-HF-01168.01-A awarded by the Space Telescope Science Institute.
We are also grateful for the repeated hospitality of
the Aspen Center for Physics and the Michigan Center for Theoretical
Physics.  

The SDSS is managed by the Astrophysical Research Consortium for the Participating Institutions. The Participating Institutions are the American Museum of Natural History, Astrophysical Institute Potsdam, University of Basel, Cambridge University, Case Western Reserve University, University of Chicago, Drexel University, Fermilab, the Institute for Advanced Study, the Japan Participation Group, Johns Hopkins University, the Joint Institute for Nuclear Astrophysics, the Kavli Institute for Particle Astrophysics and Cosmology, the Korean Scientist Group, the Chinese Academy of Sciences (LAMOST), Los Alamos National Laboratory, the Max-Planck-Institute for Astronomy (MPIA), the Max-Planck-Institute for Astrophysics (MPA), New Mexico State University, Ohio State University, University of Pittsburgh, University of Portsmouth, Princeton University, the United States Naval Observatory, and the University of Washington.





\clearpage


\clearpage

\clearpage

\clearpage

\clearpage

\clearpage

\begin{deluxetable}{clll}
\tablecolumns{4}
\tablewidth{0pc}
\tablecaption{Information available in the online cluster catalog}
\tablehead{
\colhead{Column Name} & \colhead{Data Type} & \colhead{Unit} & \colhead{Description}
}
\startdata
RA & Float & Deg & BCG Right Ascension (J2000) \\
DEC & Float & Deg & BCG Declination (J2000) \\
z & Float & & Photometric redshift \\
BCGspecz & Float & & Spectroscopic BCG redshift \\
\lrbcg & Float & $10^{10}$ \lsun & BCG r-band luminosity \\
\libcg & Float & $10^{10}$ \lsun & BCG i-band luminosity \\
\lrmem & Float & $10^{10}$ \lsun & Total r-band luminosity \\
\limem & Float & $10^{10}$ \lsun & Total i-band luminosity \\
\ngal & Int &  & Detection richness \\
\ngrtwo & Int &  & Scaled richness \\
\enddata
\tablecomments{This table describes the information provided for each cluster in the online catalog \label{total_catalog}}
\end{deluxetable}

\end{document}